\documentclass[preprint]{elsarticle}

\usepackage{natbib}
\usepackage[a4paper,margin=2.54cm]{geometry}
\usepackage{fleqn}
\usepackage{graphicx}
\usepackage[colorlinks,allcolors=blue]{hyperref}

\usepackage{placeins}
\usepackage{lineno}
\usepackage{units}
\usepackage{booktabs}
\usepackage{xspace}
\usepackage{mathtools}
\usepackage{enumitem}
\usepackage[export]{adjustbox}
\usepackage{float}
\usepackage{physics}
\usepackage{caption}
\usepackage{subcaption}

\newcommand{\wrap}[1]{\ensuremath{#1}\xspace}
\newcommand{\Mathematica}{\wrap{\textsc{Mathematica}}}
\newcommand{\GeantFour}{\wrap{\textsc{Geant4}}}
\newcommand{\oC}{\wrap{^{\circ}\textrm{C}}}
\newcommand{\GeVc}{\wrap{\textrm{GeV}/c}}

\newcommand\blfootnote[1]{%
  \begingroup
  \renewcommand\thefootnote{}\footnote{#1}%
  \addtocounter{footnote}{-1}%
  \endgroup
}

\author[a,b]{Matthew~J.~Basso}
\author[c]{Valentina~M.~M.~Cairo}
\author[d]{Chris~Damerell}
\author[e]{Dong~Su}
\author[e]{Ariel~G.~Schwartzman}
\author[e]{Jerry~Va'vra\corref{cor1}}
\cortext[cor1]{Corresponding author\ead{jjv@slac.stanford.edu}}

\affiliation[a]{organization={TRIUMF},
addressline={4004 Wesbrook Mall},
postcode={V6T 2A3},
city={Vancouver},
country={Canada}}

\affiliation[b]{organization={Department of Physics, Simon Fraser University},
addressline={8888 University Drive},
postcode={V5A 1S6},
city={Burnaby},
country={Canada}}

\affiliation[c]{organization={Experimental Physics Department, CERN},
addressline={Esplanade des Particules 1},
postcode={1211},
city={Geneva},
country={Switzerland}}

\affiliation[d]{organization={Particle Physics Department, STFC Rutherford Appleton Laboratory, Harwell Science and Innovation Campus},
postcode={OX11 0QX},
city={Didcot},
country={United Kingdom}}

\affiliation[e]{organization={SLAC National Accelerator Laboratory},
addressline={2575 Sand Hill Road},
postcode={94025-7015},
city={Menlo Park},
country={USA}}

\journal{Nucl. Instrum. Meth. A}

\begin{document}

\begin{frontmatter}

\title{A gaseous RICH detector for SiD or ILD}

\begin{abstract}

This paper describes a preliminary study of a gaseous Ring Imaging Cherenkov (RICH) system\footnote{Talk presented at the International Workshop on Future Linear Colliders (LCWS 2023), May 15--19, 2023 (C23-05-15.3).} capable of discriminating between kaons and pions at high momenta --- up to \unit[50]{\GeVc} --- and thus enhancing particle identification at future colliders. The system possesses a compact design, facilitating easy integration into existing detector concepts. A study of the key contributions to the Cherenkov angle resolution is also presented.\blfootnote{\textcopyright~2023. This manuscript version is made available under the CC-BY-NC-ND 4.0 license: \url{https://creativecommons.org/licenses/by-nc-nd/4.0/}.}

\end{abstract}


\newpageafter{title}
\newpageafter{author}
\newpageafter{abstract}

\end{frontmatter}

\section{Introduction}
\label{sec:intro}

We have made a preliminary investigation of a possible Ring Imaging Cherenkov system (RICH) detector capable of $\pi$/$K$ separation up to \unit[50]{\GeVc} at the Silicon Detector (SiD)~\cite{SiDLOI, ILCDetTDR} or International Large Detector (ILD)~\cite{ILDLOI, ILD, ILDConceptGroup:2020} at the proposed International Linear Collider (ILC)~\cite{ILCTDR}. A gaseous RICH detector is one example of a particle identification (PID) method capable of reaching such a high momentum --- see \ref{app:PID_reach}. A novel feature of our design is the use of advanced timing to reduce the silicon photomultiplier (SiPM) noise.

This effort is part of larger physics study~\cite{Albert:2022mpk} arguing the importance of this type of detector for a future linear collider.

\section{Overall concept}

\begin{figure}[htbp]
    \centering
    \begin{subfigure}{0.8\textwidth}
        \centering
        \includegraphics[width=1.\textwidth]{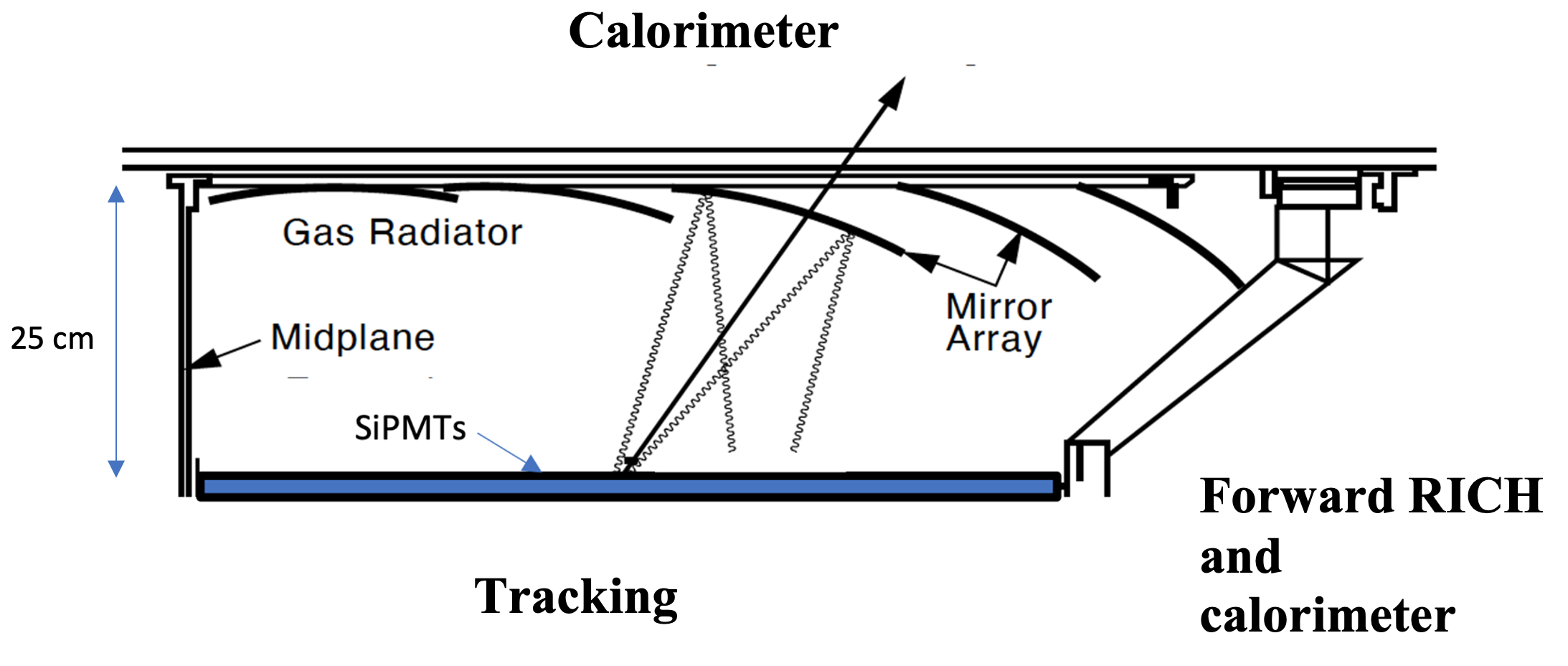}
        \caption{Side view of overall layout}
    \end{subfigure} \\
    \begin{subfigure}[b]{0.65\textwidth}
        \centering
        \includegraphics[width=1.\textwidth]{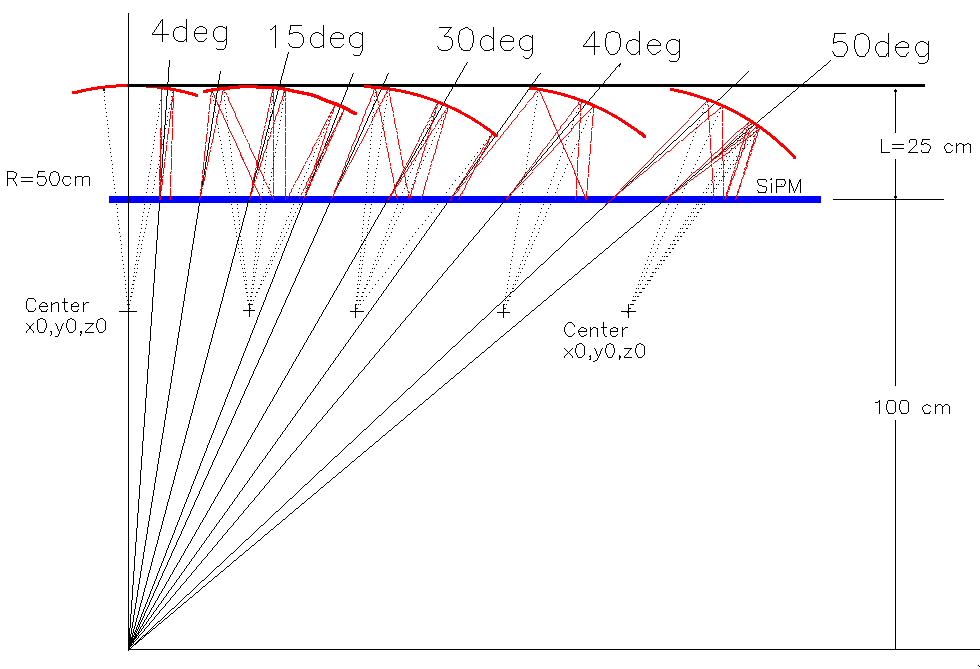}
        \caption{Side view with tracks}
        \label{fig:proposed_RICH_side}
    \end{subfigure}
    \hfill
    \begin{subfigure}[b]{0.30\textwidth}
        \centering
        \includegraphics[width=1.\textwidth]{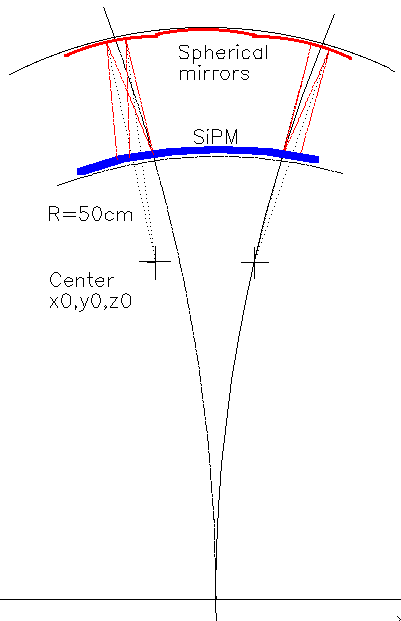}
        \caption{Front view with tracks}
        \label{fig:proposed_RICH_front}
    \end{subfigure}
    \caption{Proposed gaseous RICH detector at SiD/ILD. (a) Side view showing the relative placement of the tracking, calorimetry, and forward instrumentation is indicated. Ray-tracing via the \textsc{Graphite} program is used to define the positions of the mirrors. (b) Side view and (c) front view of the proposed detector with tracks. All of the mirrors have a radius of \unit[50]{cm}. This optical design is preliminary as further tuning of the mirror positions is required.}
    \label{fig:proposed_RICH}
\end{figure}

The optical concept of the detector is shown in Fig.~\ref{fig:proposed_RICH}. It includes full SiPM coverage so that one may use the time difference between the track and Cherenkov photon hits in order to reduce the SiPM noise; the same result could be achieved using a special timing layer providing the track start time, with the SiPM providing the photon stop signal. Our initial choice for the RICH detector's thickness is \unit[25]{cm} active radial length. This could be optimized if a better photon detector becomes available.

The RICH detector uses spherical mirrors and SiPM photon detectors whose coverage follows the shape of the barrel of a cylinder. Fig.~\ref{fig:proposed_RICH} resembles the gaseous RICH detector of the SLAC Large Detector's (SLD's) Cherenkov Ring Imaging Detector (CRID)~\cite{Vavra:1999}; however, introducing a SiPM-based design improves the PID performance significantly compared to SLD's and DELPHI's gaseous RICH detectors~\cite{Albrecht:1999ri}.

Although we have selected a specific type of SiPM in this paper, we believe that photon technology will improve over the next 15~years in terms of noise performance, timing capability, pixel size, and detection efficiency. The overall aim is to make this RICH detector with as low mass as possible because we do not want to degrade the calorimeter. This speaks for mirrors made of beryllium~\cite{Barber:2006, JWST} and a structure made of low mass carbon-composite material. Another important aspect is to make the RICH detector depth as thin as possible in order to reduce the cost of the calorimeter. Our initial choice of \unit[25]{cm} could be reduced further if the detection efficiency of photon detectors improves.

\subsection{Gas choices}

We have considered several different gases\footnote{All of the gases considered are fluorocarbons, which may prove difficult to source on the timescale of the future collider due to their environmental impact; however, the environmental impact may be limited by capturing the gas, cleaning it, and recirculating it for use.} as a radiator, including:

\begin{enumerate}[label=(\alph*)]
    \item Pure C$_5$F$_{12}$ gas at \unit[1]{bar} requires a detector temperature of \unit[+40]{\oC} since the boiling point of this gas is \unit[+31]{\oC} at \unit[1]{bar}. That could prove to be difficult since SiPMs need to be cooled;
    \item A gas choice of pure C$_4$F$_{10}$ at \unit[1]{bar} allows detector operation at a few degrees Celsius since boiling point of this gas is \unit[$-$1.9]{\oC} at \unit[1]{bar}. This is presently our \emph{preferred} choice;
    \item A choice of C$_2$F$_6$ gas at \unit[1]{bar} would allow detector operation even below \unit[0]{\oC} since the boiling point of this gas is \unit[$-$70.2]{\oC} at \unit[1]{bar}. However, this gas would deliver an insufficient number of photoelectrons for the geometry shown in Fig.~\ref{fig:proposed_RICH} and therefore it was not considered;
    \item A choice of C$_3$F$_8$ gas at \unit[1]{bar} would allow detector operation at \unit[$-$30]{\oC} since the boiling point of C$_3$F$_8$ is \unit[$-$37]{\oC}. The detector's PID performance will be between that of C$_2$F$_6$ and C$_4$F$_{10}$. It is certainly worthwhile to look into this solution.
\end{enumerate}

\subsection{Number of photoelectrons per ring}

\begin{figure}[htbp]
    \centering
    \begin{subfigure}[b]{0.44\textwidth}
        \centering
        \includegraphics[width=1.\textwidth,valign=b]{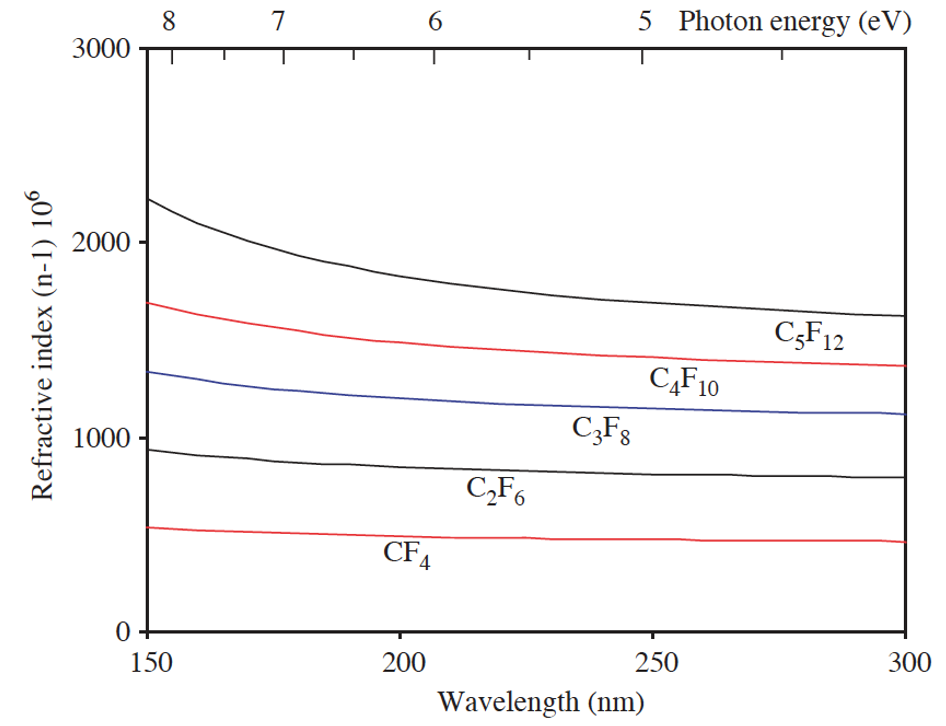}
        \caption{}
        \label{fig:efficiency_refractive_index}
    \end{subfigure}
    \hfill
    \begin{subfigure}[b]{0.54\textwidth}
        \centering
        \includegraphics[width=1.\textwidth,valign=b]{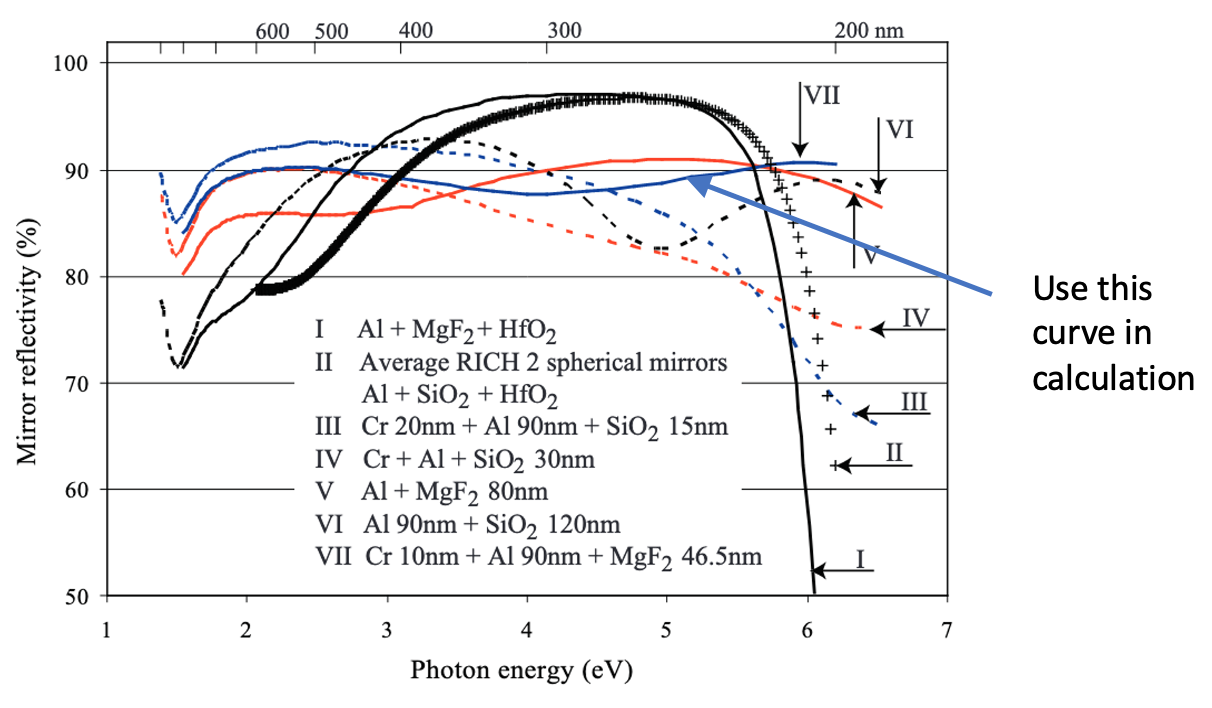}
        \caption{}
        \label{fig:efficiency_reflectivity}
    \end{subfigure} \\
    \vspace{0.5em}
    \begin{subfigure}[b]{0.54\textwidth}
        \centering
        \includegraphics[width=1.\textwidth,valign=b]{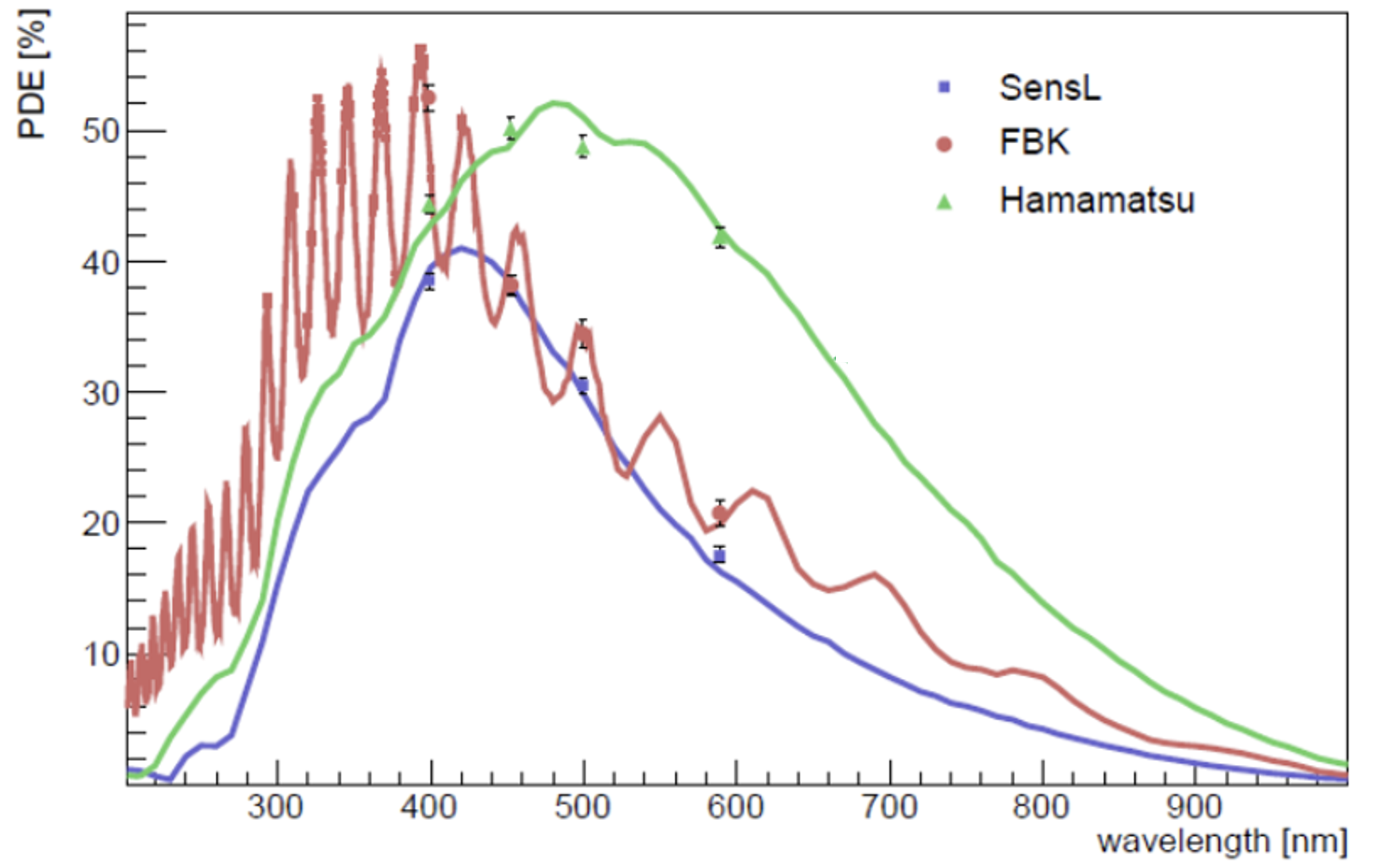}
        \caption{}
        \label{fig:efficiency_PDE_1}
    \end{subfigure}
    \hfill
    \begin{subfigure}[b]{0.44\textwidth}
        \centering
        \includegraphics[width=1.\textwidth,valign=b]{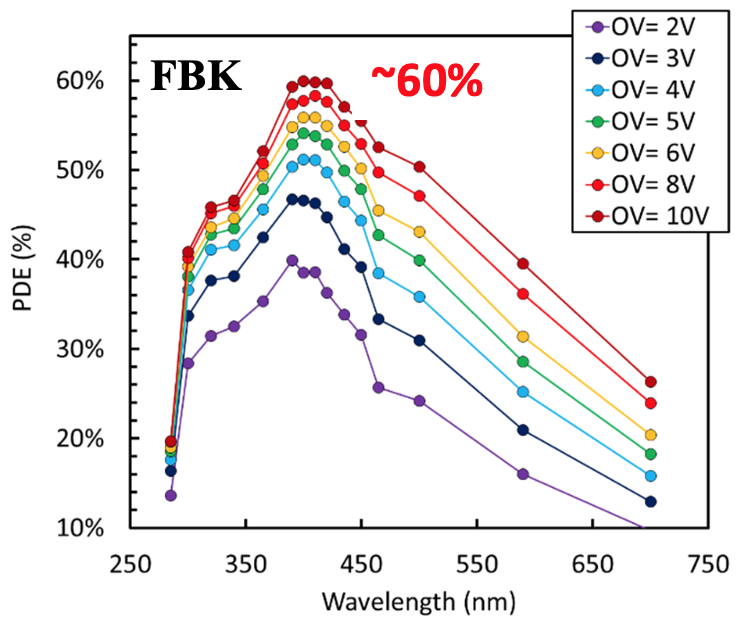}
        \caption{}
        \label{fig:efficiency_PDE_2}
    \end{subfigure} \\
    \vspace{0.5em}
    \begin{subfigure}[b]{0.41\textwidth}
        \centering
        \includegraphics[width=1.\textwidth,valign=b]{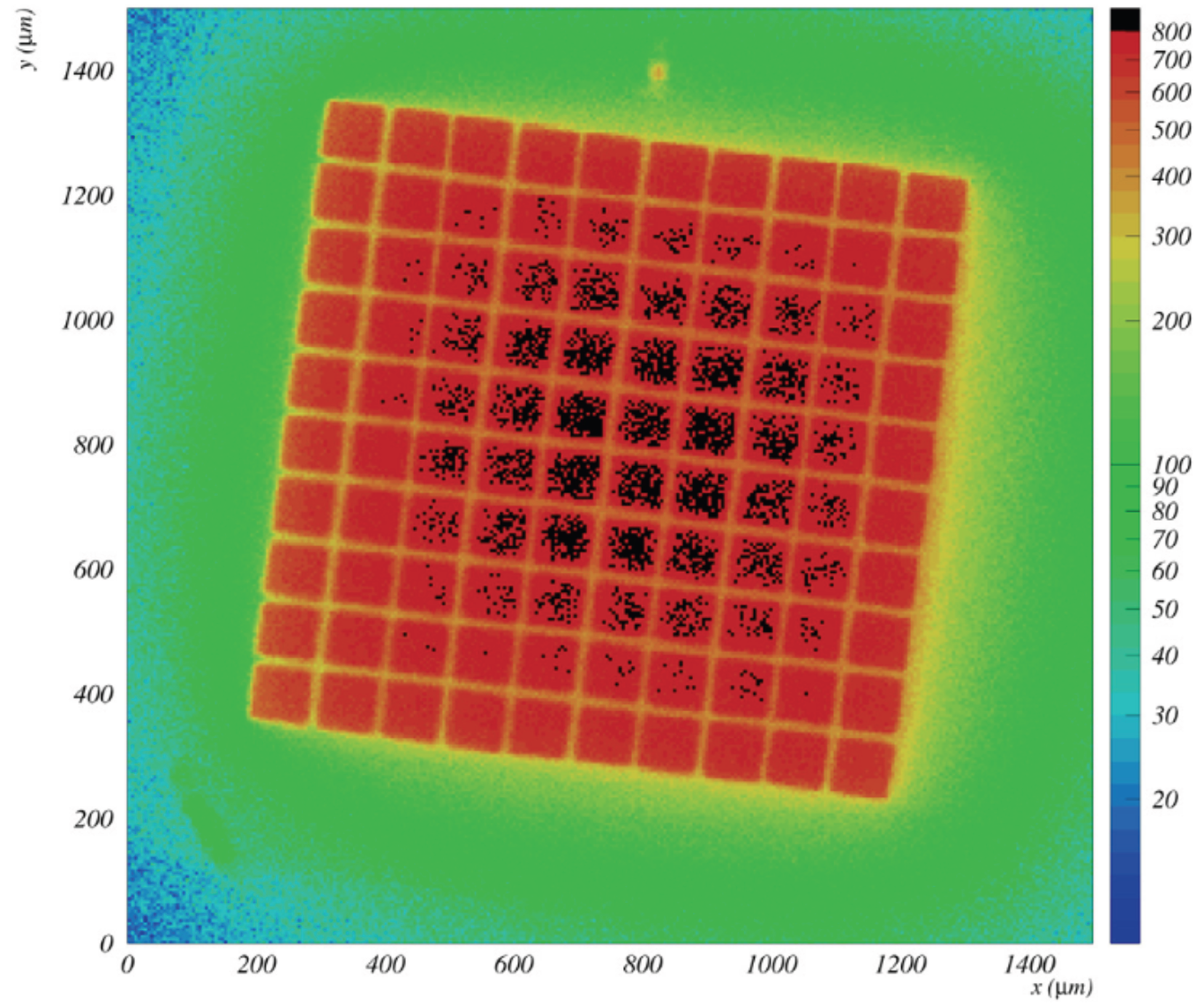}
        \caption{}
        \label{fig:efficiency_PMT}
    \end{subfigure}
    \hfill
    \begin{subfigure}[b]{0.52\textwidth}
        \centering
        \includegraphics[width=1.\textwidth,valign=b]{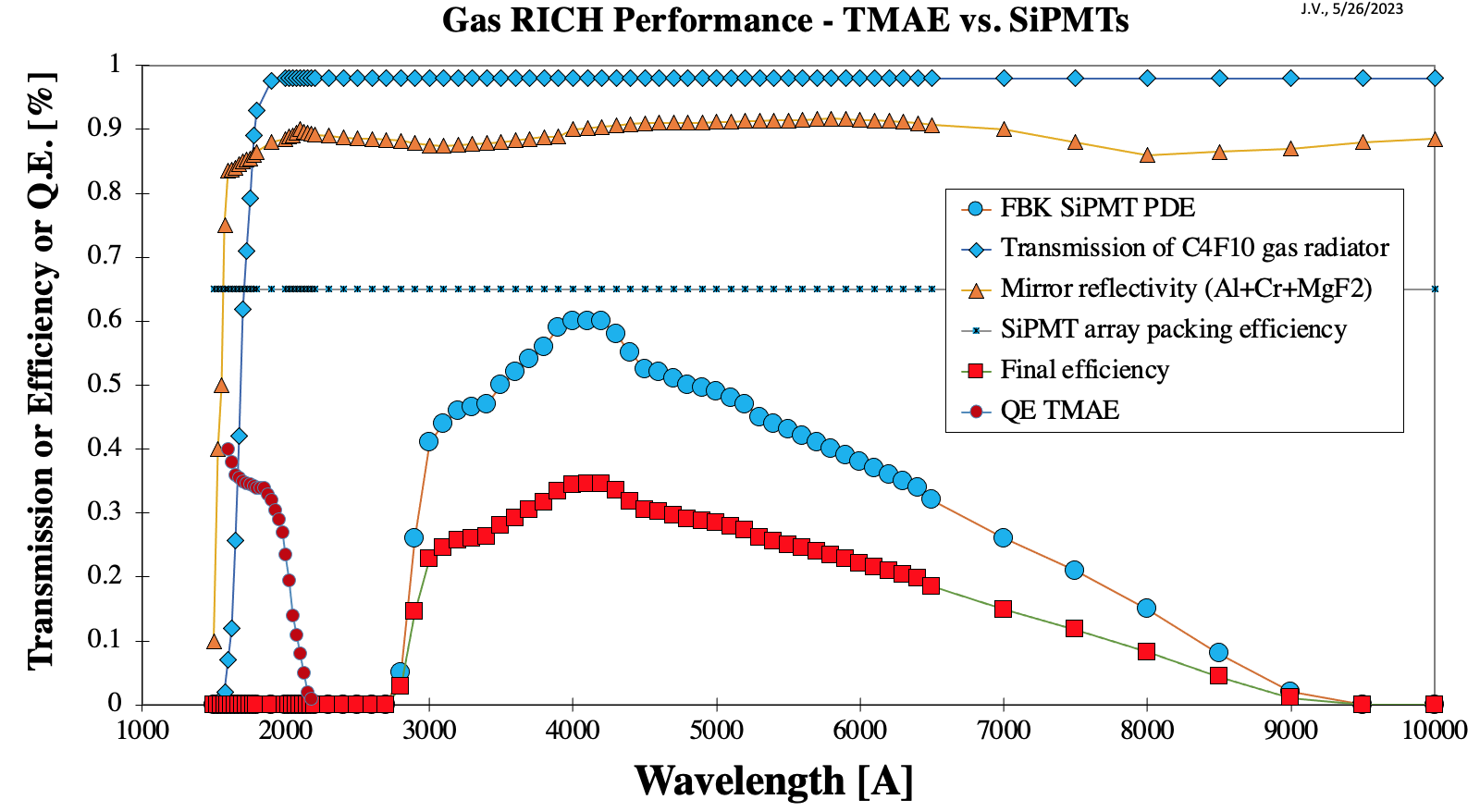}
        \caption{}
        \label{fig:efficiency_separate}
    \end{subfigure}
    \caption{(a) Refractive index for the gases considered~\cite{Vavra:2014, Ullaland:2005}. (b) Reflectivity of various mirror coatings~\cite{LHCb:2008}; we used Cr/Al/MgF$_2$ coating in our calculation. (c) Photon detection efficiency (PDE) of a single SiPM from several sources~\cite{NepomukOtte:2016}; we used the Hamamatsu curve in our calculation. (d) PDE of the FBK SiPM (60\%) used in our calculation~\cite{GolaConf} as a comparison with the Hamamatsu curve in (c). (e) A SiPM array has additional losses due to gaps between pixel elements~\cite{Korpar:2020}, the so called ``packing efficiency''. (f) The various efficiencies, including packing efficiency, gas transmission, mirror reflectivity, and the FBK SiPM PDE, used in our calculation.}
    \label{fig:efficiency}
\end{figure}

\begin{figure}[htbp]
    \centering
    \includegraphics[width=0.8\textwidth]{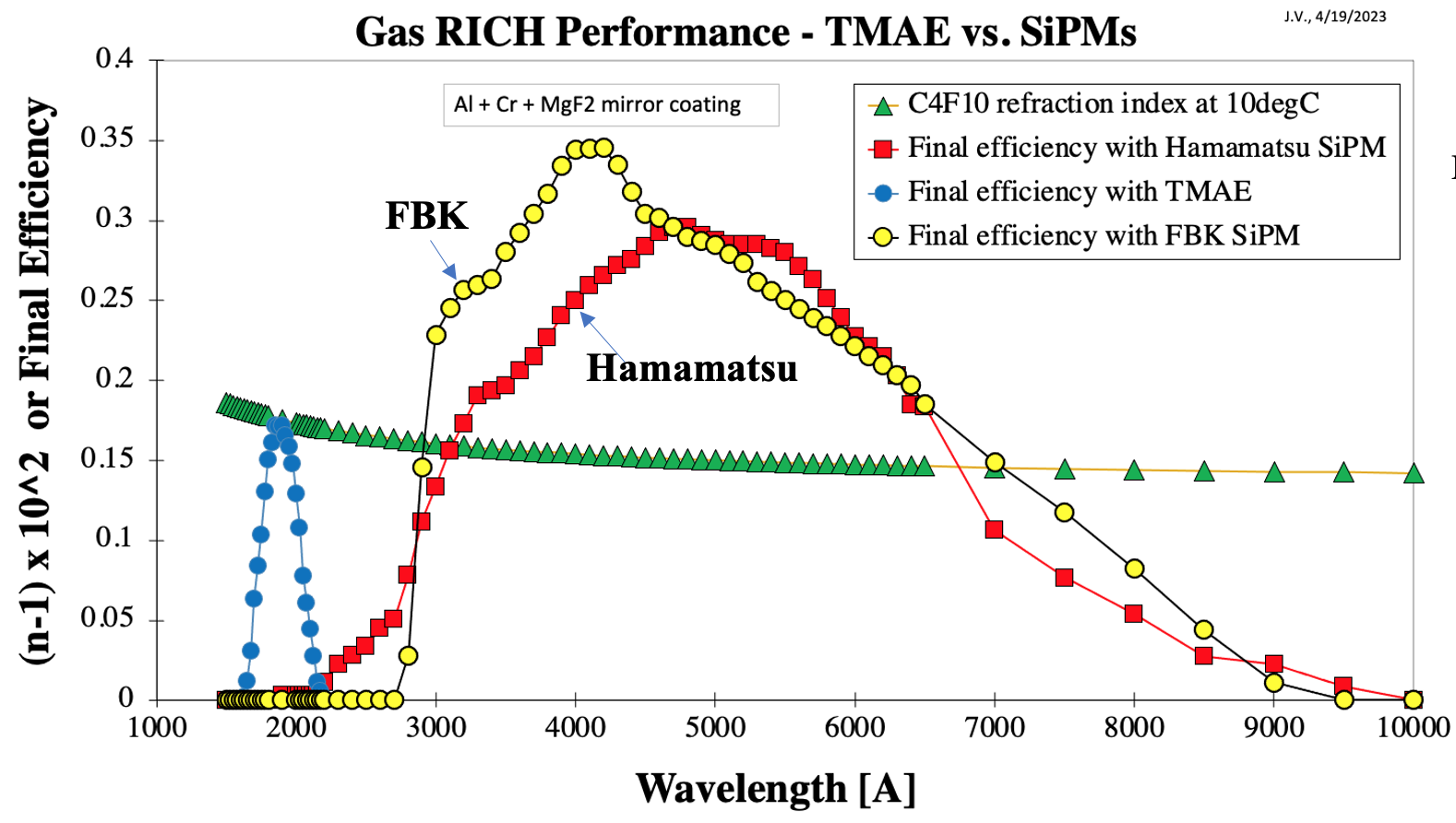}
    \caption{Final efficiency of the SiPM compared to the final efficiency of the SLD CRID with the TMAE photocathode, as calculated in this work. The refractive index of C$_4$F$_{10}$ is also plotted to indicate the chromaticity in our detector proposal.}
    \label{fig:efficiency_combined}
\end{figure}

The number of photoelectrons, $N_\textrm{pe}$, is calculated using:

\begin{equation}
    N_\textrm{pe} = N_0 L \sin^2\!\left(\langle\theta_c\rangle\right) \,,
\end{equation}

\noindent where $L$ is the length of the radiator, $\langle\theta_c\rangle$ is the mean Cherenkov angle, and:

\begin{equation}
    N_0 = \frac{\alpha}{\hbar c}\frac{\int\varepsilon(E)\sin^2\!\theta_c\dd{E}}{\sin^2\!\left(\langle\theta_c\rangle\right)} \,\,\,\,\textrm{with}\,\,\,\, \frac{\alpha}{\hbar c} = \unit[370.5]{eV^{-1}cm^{-1}} \,,
\end{equation}

\noindent where $\alpha$ is the fine-structure constant, $\hbar$ is the reduced Planck's constant, $c$ is the speed of light, and $E$ is the energy of the photon. The Cherenkov angle, $\theta_c$, is given by:

\begin{equation}
    \cos\theta_c(\lambda) = \frac{1}{n(\lambda)\beta} \,,
\end{equation}

\noindent where $\lambda$ is the wavelength of the photon, $n$ is the refractive index of the medium, and $\beta = v/c$. To calculate $N_0$, one also needs to calculate $\varepsilon(E)$, which is the product of all of the efficiencies in the problem, and to determine the refractive index as a function of wavelength to calculate the Cherenkov angle. Fig.~\ref{fig:efficiency_refractive_index} shows the refractive index for all gases considered~\cite{Vavra:2014, Ullaland:2005}. Fig.~\ref{fig:efficiency_reflectivity} shows reflectivity of various mirror coatings~\cite{LHCb:2008}. We chose the reflectivity of Cr/Al/MgF$_2$ coating in the calculation, as indicated on the graph, although a Al/Cr/HfO$_2$ coating could also be considered in future. Figs.~\ref{fig:efficiency_PDE_1} and \ref{fig:efficiency_PDE_2} show photon detection efficiency (PDE) of a single SiPM~\cite{NepomukOtte:2016, GolaConf}. We have chosen the FBK PDE for our calculation. Fig.~\ref{fig:efficiency_PMT} shows that a SiPM array has additional losses due to gaps between the pixel elements of the array~\cite{Korpar:2020}, the so called ``packing efficiency''. We have chosen a packing efficiency of 65\% in our calculation. Fig.~\ref{fig:efficiency_separate} shows the various efficiencies used in our calculation, and Fig.~\ref{fig:efficiency_combined} shows the final efficiency of the SiPM-based and the TMAE\footnote{``TMAE'' $\coloneqq$ ``tetrakis(dimethylamine)ethylene''.}-based detector solutions used by the SLD CRID and the DELPHI RICH. The SiPM solution is vastly better than the TMAE solution in terms of overall efficiency, as one can see from Fig.~\ref{fig:efficiency_combined}.\footnote{However, because the number of Cherenkov photons is proportional to $1/\lambda^2$, the TMAE-based RICH is considerably better than Figs.~\ref{fig:efficiency_separate} and \ref{fig:efficiency_combined} would suggest.}

\begin{figure}[htbp]
    \centering
    \begin{subfigure}[b]{0.49\textwidth}
        \centering
        \includegraphics[width=1.\textwidth]{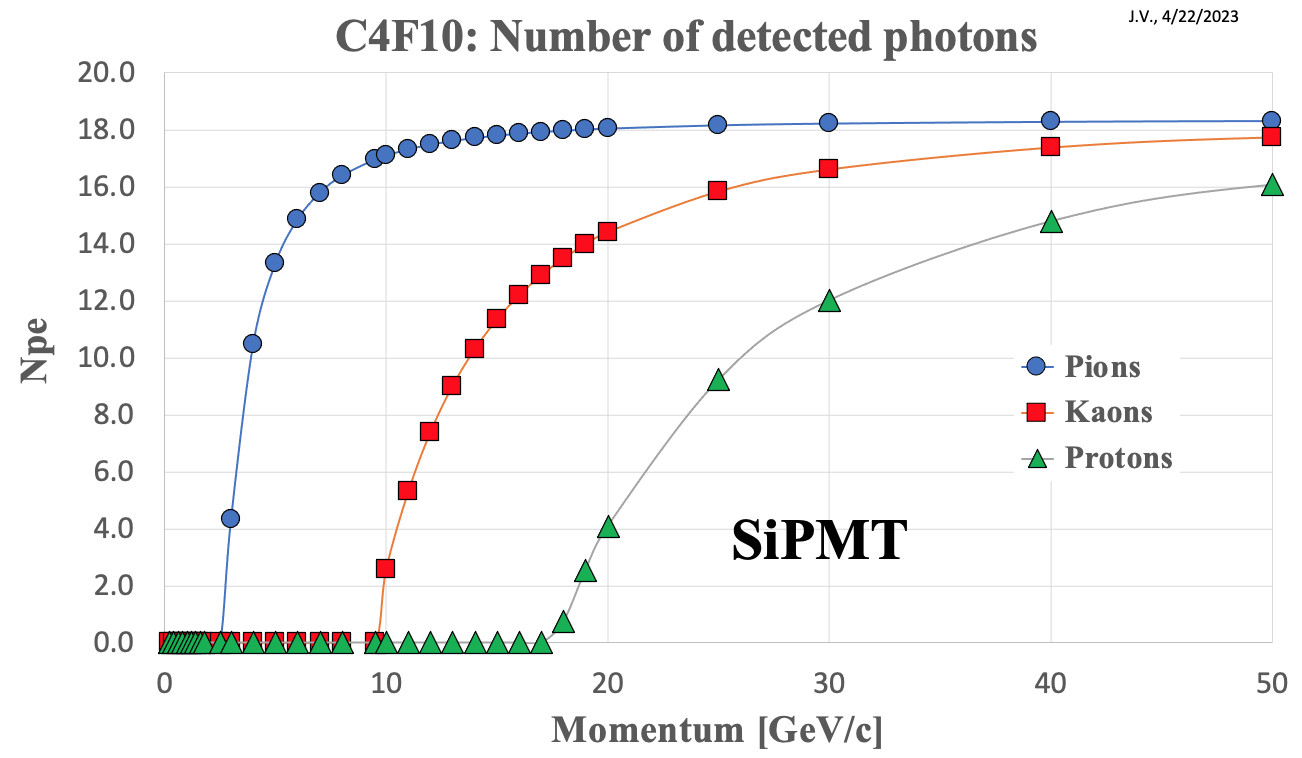}
        \caption{}
    \end{subfigure}
    \hfill
    \begin{subfigure}[b]{0.49\textwidth}
        \centering
        \includegraphics[width=1.\textwidth]{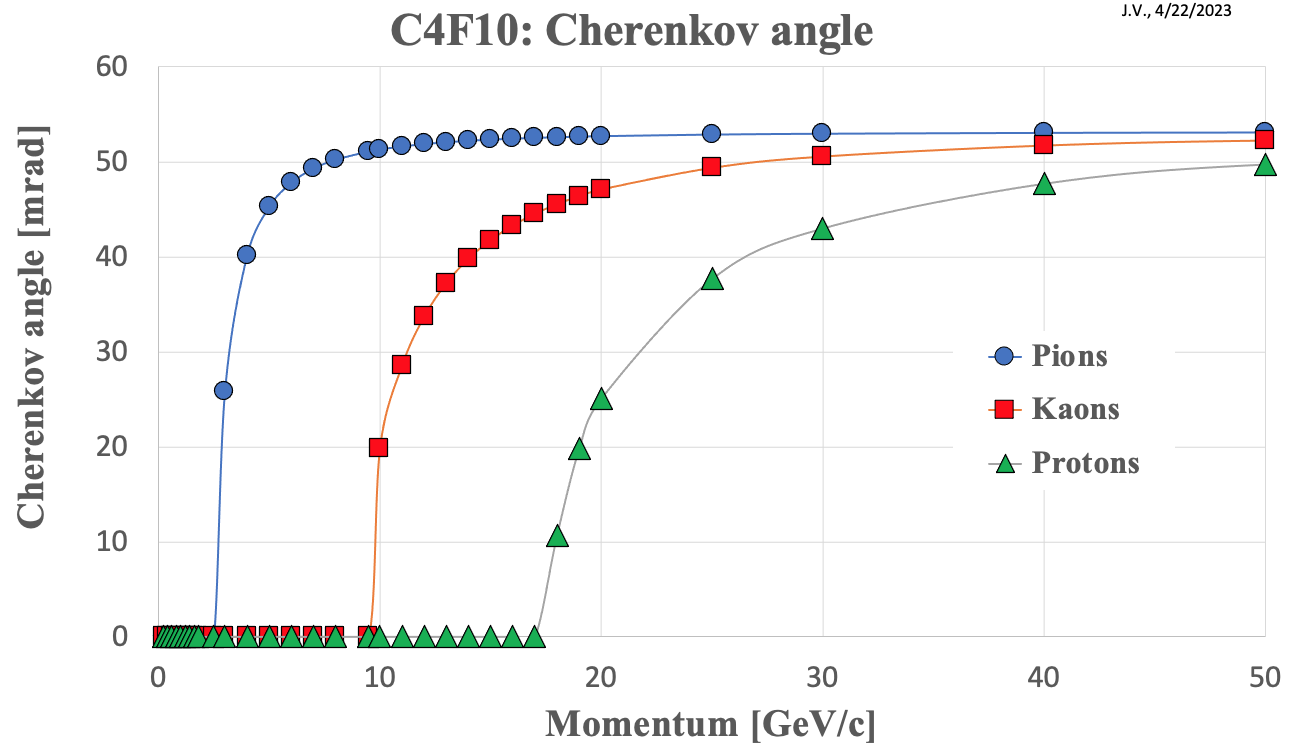}
        \caption{}
    \end{subfigure} \\
    \caption{(a) Calculated number of photoelectrons per ring and (b) Cherenkov angle, each as a function of momentum for pions, kaons, and protons.}
    \label{fig:number_of_photoelectrons}
\end{figure}

Fig.~\ref{fig:number_of_photoelectrons} shows the calculated number of photoelectrons per ring as well as the Cherenkov angle, each as a function of momentum. One can see that the kaon threshold is at \unit[$\sim$10]{\GeVc} for C$_4$F$_{10}$ gas and that the expected number of photoelectrons per ring is about 18 for $L = \unit[25]{cm}$ and $\beta \sim 1$. For comparison, the SLD CRID's gaseous RICH had $\sim$10 photoelectrons per ring for an \unit[80\%]{C$_5$F$_{12}$}\,/\,\unit[20\%]{N$_2$} mix and $L = \unit[45]{cm}$ and $\beta \sim 1$~\cite{Vavra:1999}.

\subsection{PID performance as a function of Cherenkov angle resolution}

\begin{figure}
    \centering
    \includegraphics[width=0.8\textwidth]{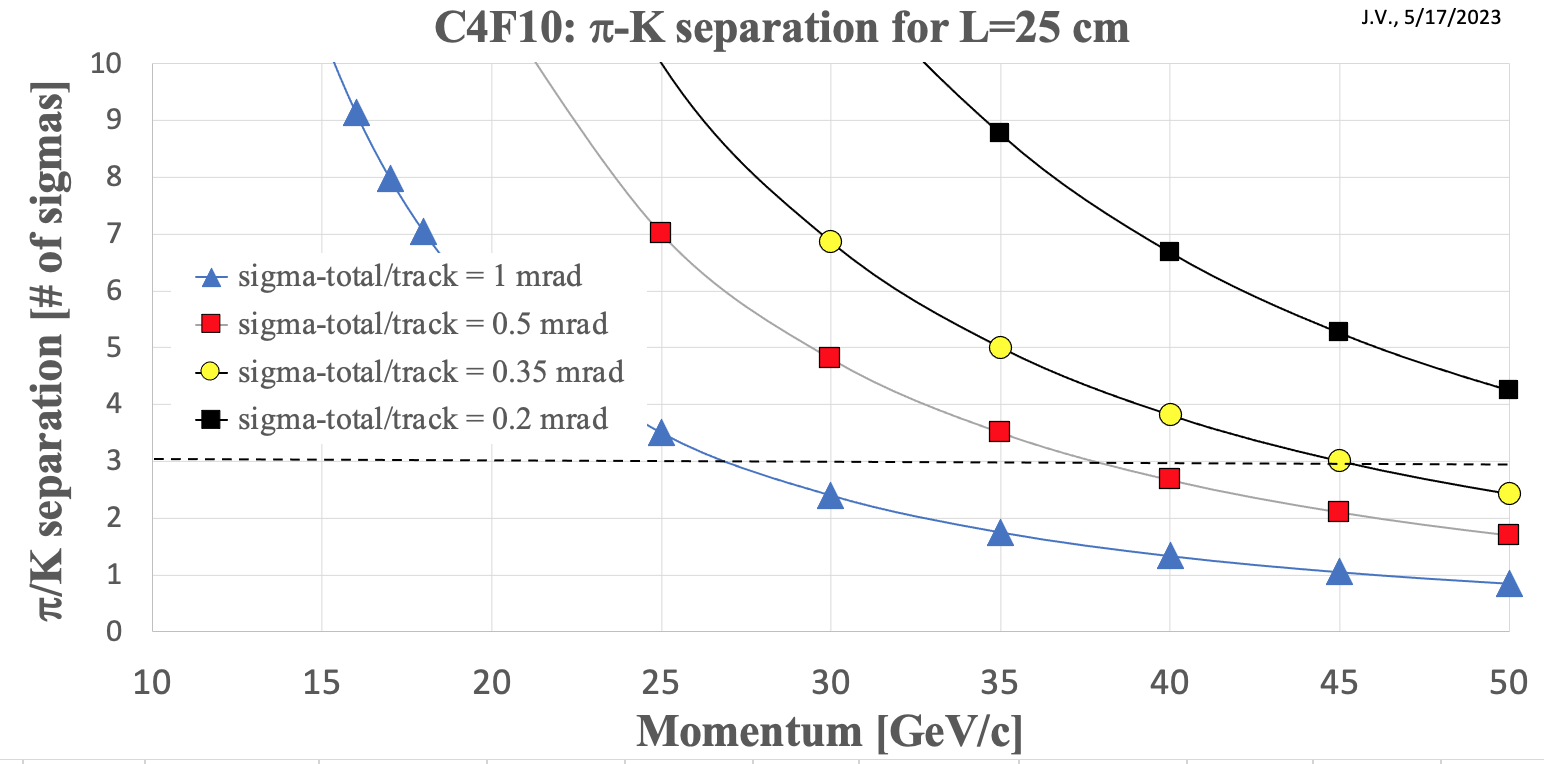}
    \caption{Expected PID performance as a function of momentum and total Cherenkov angle resolution per track.}
    \label{fig:expected_sigmas}
\end{figure}

The RICH detector performance can be divided into a threshold region, where one can identify particles based on threshold, ring size, and number of photoelectrons per ring (see Fig.~\ref{fig:number_of_photoelectrons}), and a high momentum region, where one can use the following formula to determine the particle separation $S$ (in number of sigmas):

\begin{equation}
    S = \frac{|\theta_\pi - \theta_K|}{(\sigma_{\theta_c} / \sqrt{N_\textrm{pe}})} \,,
\end{equation}

\noindent where $\theta_\pi$ is the Cherenkov angle for pions, $\theta_K$ is the Cherenkov angle for kaons, $\sigma_{\theta_c}$ is the single-photon Cherenkov angle resolution, and $N_\textrm{pe}$ is number of photoelectrons per ring. Fig.~\ref{fig:expected_sigmas} shows the PID performance of the proposed detector for a C$_4$F$_{10}$ gas as a function of the total Cherenkov angle resolution per track, given by the Cherenkov angle resolution per photon divided by $\sqrt{N_\textrm{pe}}$ and added in quadrature to the tracking resolution. SLD's and DELPHI's gaseous RICH detectors had resolutions of \unit[$\sim$1]{mrad} per track, limiting their PID reach to \unit[25--30]{\GeVc}. To achieve PID performance up to \unit[$\sim$50]{\GeVc}, it is essential to limit this resolution to less than \unit[0.3]{mrad} per track.

\section{Resolution contributions to the Cherenkov angle measurement}

In this section, we will discuss the various contributions to the Cherenkov angle resolution. There are contributions per single photon, which get divided by $\sqrt{N_\textrm{pe}}$. Examples of such contributions are the chromatic, pixel, and smearing/focusing errors, and their total contribution is given by their sum in quadrature:

\begin{equation}
    \sigma_\textrm{single-photon} = \sqrt{(\sigma_\textrm{chromatic}/\textrm{photon})^2 + (\sigma_\textrm{pixel}/\textrm{photon})^2 + (\sigma_\textrm{smearing/focusing}/\textrm{photon})^2} \,.
    \label{eqn:sigma_single_photon}
\end{equation}

\noindent In addition, there are overall errors, which do not get reduced by $\sqrt{N_\textrm{pe}}$. Examples include the tracking error and other contributions from correlated terms such as alignment, multiple scattering, hit ambiguities, background hits from random sources, hits from other tracks, SiPM noise hits, and cross talk. The total Cherenkov error per track in the general case is given by:

\begin{equation}
    \sigma_{\theta_c}/\textrm{track} = \sqrt{(\sigma_\textrm{single-photon}/\sqrt{N_\textrm{pe}})^2 + \sigma_\textrm{tracking}^2 + \sigma_\textrm{other}^2} \,.
    \label{eqn:sigma_total}
\end{equation}

\noindent To reach $\pi$/$K$ separation at the $\sim$3$\sigma$ level, it is necessary to keep $\sigma_{\theta_c}/\textrm{track}$ at a level of \unit[$\leq 0.3$]{mrad}. This is not a trivial task, as it requires an exceptional tracking angular resolution in the range of \unit[0.1--0.2]{mrad} as well as keeping all systematic errors below \unit[$\sim$0.2]{mrad}.

\subsection{Chromatic error}

\begin{figure}[htbp]
    \centering
    \includegraphics[width=0.8\textwidth]{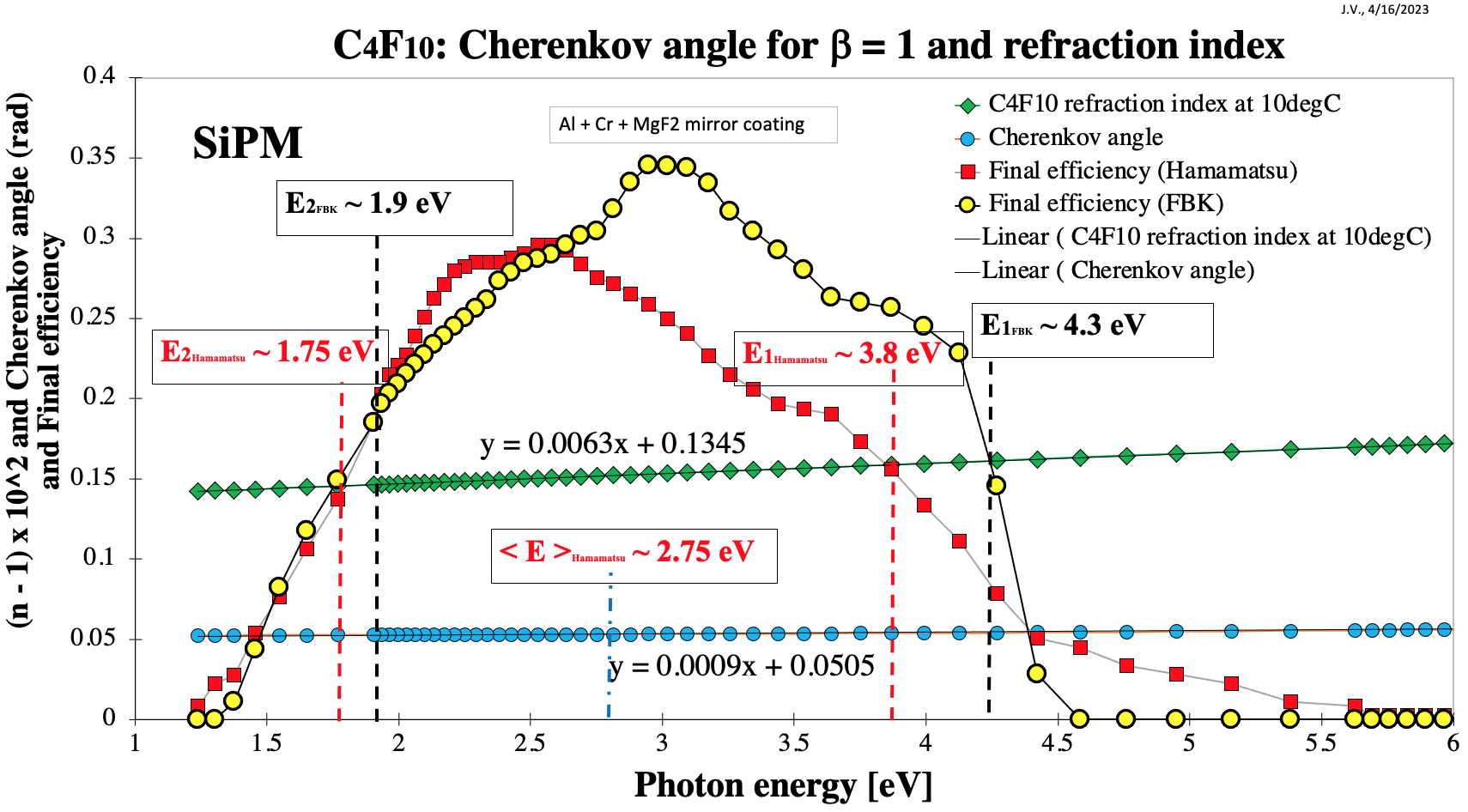}
    \caption{Final efficiency for the FBK SiPM design. The refractive index of C$_4$F$_{10}$ is also plotted to indicate the chromaticity in our detector proposal.}
    \label{fig:final_efficiency}
\end{figure}

The chromatic error may affect the RICH performance significantly. Although the SLD CRID, using TMAE, operated in a region where the refractive index changed more rapidly, its wavelength acceptance was very narrow and therefore the chromatic error was smaller than that of a SiPM-based detector. From Figs.~\ref{fig:efficiency_combined} and \ref{fig:final_efficiency}, we determine the average wavelength to be \unit[450]{nm} (or \unit[2.75]{eV}), which corresponds to an average refractive index of $n \sim 1.001434$. For SiD/ILD, we determine from Fig.~\ref{fig:final_efficiency} that the chromatic error contribution for our RICH is $\sigma_{\theta_c} \sim (d\theta_c/dE)(E_2 - E_1)/\sqrt{12} \sim \unit[0.62]{mrad/photon}$, which is larger than that of the SLD CRID, \unit[$\sim$0.4]{mrad/photon}, determined using the same method. This large chromatic error is due to a very broad wavelength acceptance provided by the SiPM-based design.

\subsection{Error due to a finite SiPM pixel size}

We assume that SiPMs will have $\unit[0.5]{mm}\times\unit[0.5]{mm}$ pixels. The Cherenkov error contribution due to the finite pixel size is $\sigma_{\theta_c} \sim (\unit[0.05]{cm} / \sqrt{12}) / (1.5 \times \unit[25]{cm}) \sim \unit[0.38]{mrad/photon}$. In this simple calculation, we assume that the Cherenkov photons are produced on average halfway through the radiator of thickness \unit[25]{cm}. Therefore, we assume a total photon path length of $\unit[25/2]{cm} + \unit[25]{cm} = \unit[1.5 \times 25]{cm}$ to determine the pixel angular error.

\subsection{Cherenkov angle focusing and smearing error}
\label{sec:focusing_and_smearing_error}

\begin{figure}[htbp]
    \centering
    \includegraphics[width=\textwidth]{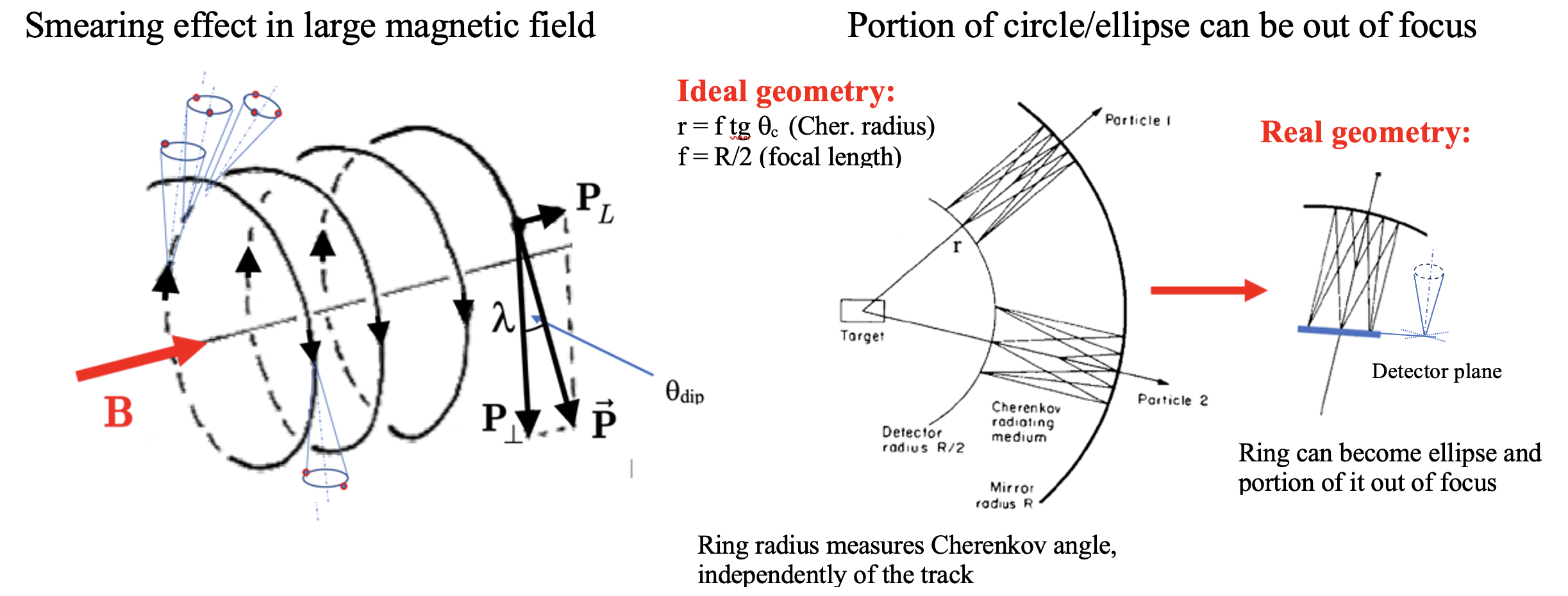}
    \caption{(Left) schematic diagram of the helix trajectory and Cherenkov cones. (Right) ideal and real spherical geometries. In the ideal geometry, the Cherenkov ring is in perfect focus; in a real geometry, the ring can become an ellipse and a portion of it can be out-of-focus --- see Fig.~\ref{fig:ring_distortions}.}
    \label{fig:helix_trajectory}
\end{figure}

Running this type of RICH detector at \unit[5]{T} has some consequences: there is a considerable contribution to the Cherenkov angle error due to a magnetic field smearing effect for tracks with momenta below \unit[20]{\GeVc}. Fig.~\ref{fig:helix_trajectory} shows that the Cherenkov cone rotates in 3D as the particle trajectory follows a helix. This contributes to the smearing of the image, and it affects the detected points around the Cherenkov azimuth angle $\phi_c$ differently and is generally larger for larger magnetic fields $B$, larger dip angles\footnote{The dip angle is defined relative to the axis transverse to the beam.} $\theta_\textrm{dip}$, and smaller momenta $p$.

\begin{figure}[htbp]
    \centering
    \includegraphics[width=\textwidth]{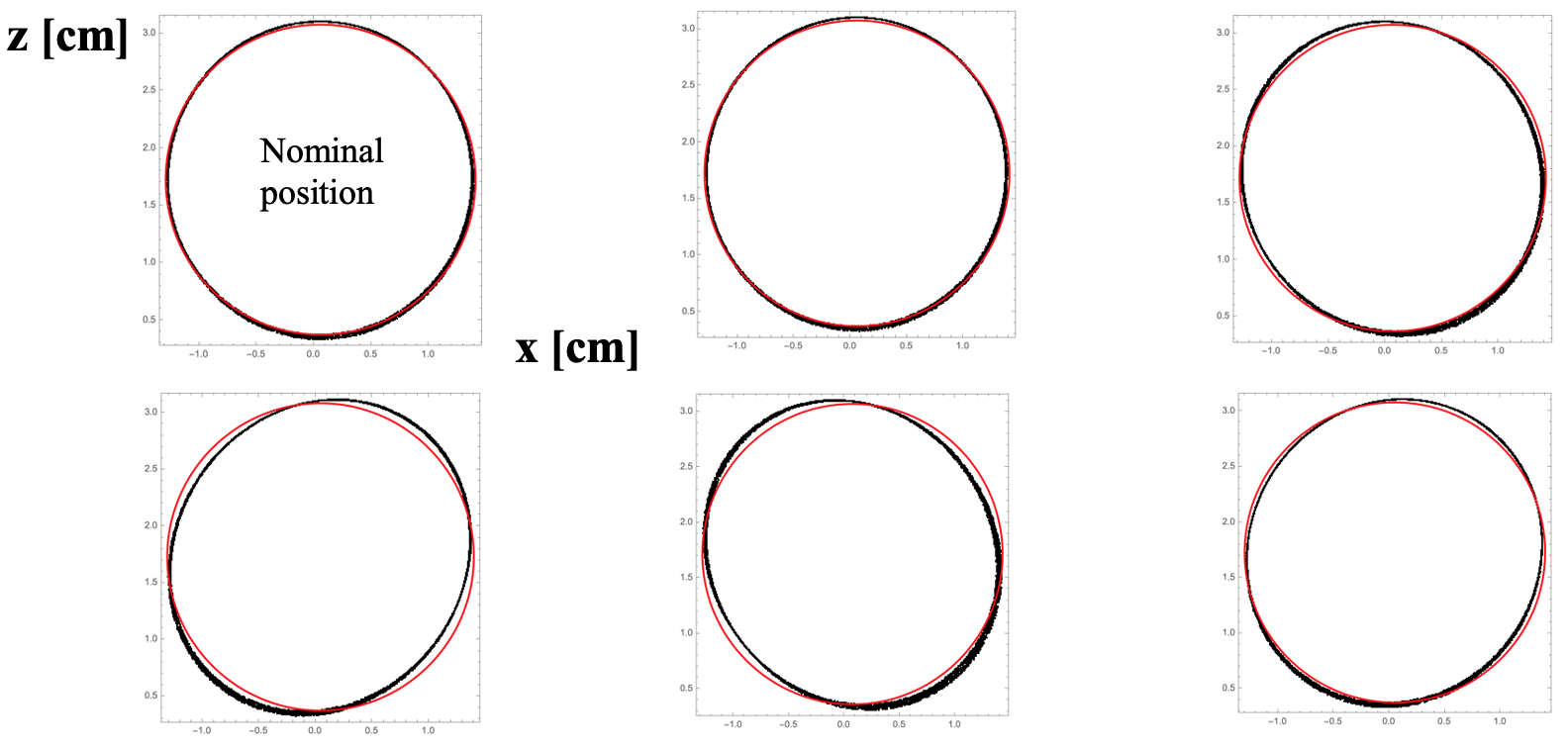}
    \caption{Illustrations of ring distortions at $\theta_\textrm{dip} = 4^\circ$, $p = \unit[20]{\GeVc}$, and $B = \unit[5]{T}$ using a \Mathematica code --- see Section~\ref{sec:focusing_and_smearing_error}. The detector plane was rotated around the perpendicular axis of the nominal detector position --- see the rightmost diagram of Fig.~\ref{fig:helix_trajectory}.}
    \label{fig:ring_distortions}
\end{figure}

In addition, we have an error due to the focusing of the ring. Fig.~\ref{fig:helix_trajectory} also shows the ideal detector geometry for a cylindrical mirror of radius $R$; an ideal detector surface should follow a sphere of radius $R/2$. In reality, a real detector geometry does not follow a spherical surface, which means that a portion of the ring is out-of-focus --- this is also clear in Figs.~\ref{fig:proposed_RICH_side} and \ref{fig:proposed_RICH_front}. As for the smearing effect, the focusing effect is also dependent on the Cherenkov azimuth angle. In the following paragraphs, we will try to estimate both effects. The focusing effect can be minimized by detector plane rotations; however, this has to be done by simulating all possible track directions and momenta. In this paper, we will assume that the detectors follow the shape of a cylinder, as shown in Figs.~\ref{fig:proposed_RICH_side} and \ref{fig:proposed_RICH_front}, and that each detector is planar (probably $\unit[10]{cm}\times\unit[10]{cm}$). Fig.~\ref{fig:ring_distortions} illustrates ring distortions at $\theta_\textrm{dip} = 4^\circ$, $p = \unit[20]{\GeVc}$, and $B = \unit[5]{T}$ using a \Mathematica code. The detector plane was rotated around the perpendicular axis of the nominal detector position. The images are ellipses and the out-of-focus portions are changing for different orientations of the planar detector.

\begin{figure}[htbp]
    \centering
    \includegraphics[width=\textwidth]{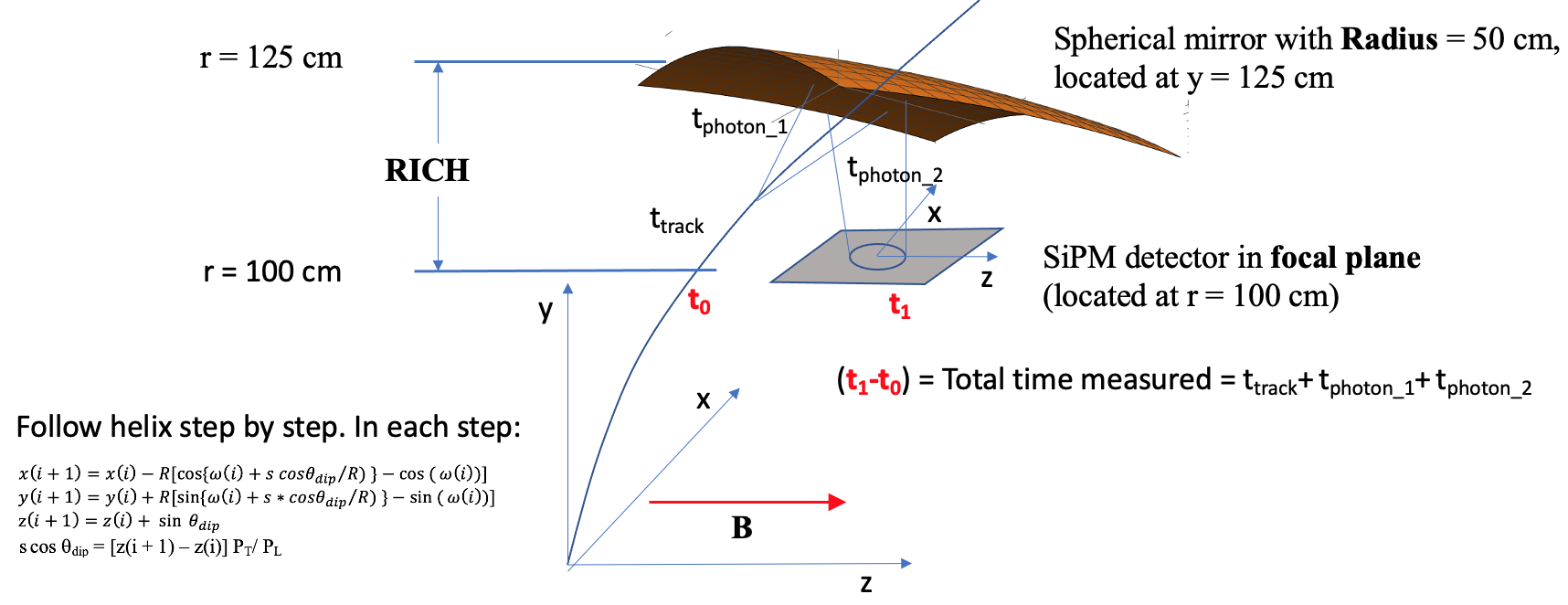}
    \caption{A schematic diagram of the helix trajectory and Cherenkov light. A simple 3D model was implemented in \Mathematica: step through the magnetic field, radiate Cherenkov photons when $100 < r < \unit[125]{cm}$, reflect them from a spherical mirror, and find their intersection with a detector plane. The input for the positions of the mirrors came from the ray tracing program, schematically shown in Figs.~\ref{fig:proposed_RICH_side} and \ref{fig:proposed_RICH_front}.}
    \label{fig:simulation}
\end{figure}

The \Mathematica code steps charged particles in a magnetic field following a helix. Fig.~\ref{fig:simulation} shows schematically the simulation model. Once in the radiator region ($100 < r < \unit[125]{cm}$), particles radiate Cherenkov photons. Photons reflect from a spherical mirror and are imaged on a plane of SiPMs. We will discuss in this paper only the case where the SiPM detector plane is horizontal at $y = \unit[100]{cm}$.

\begin{figure}[htbp]
    \centering
    \includegraphics[width=\textwidth]{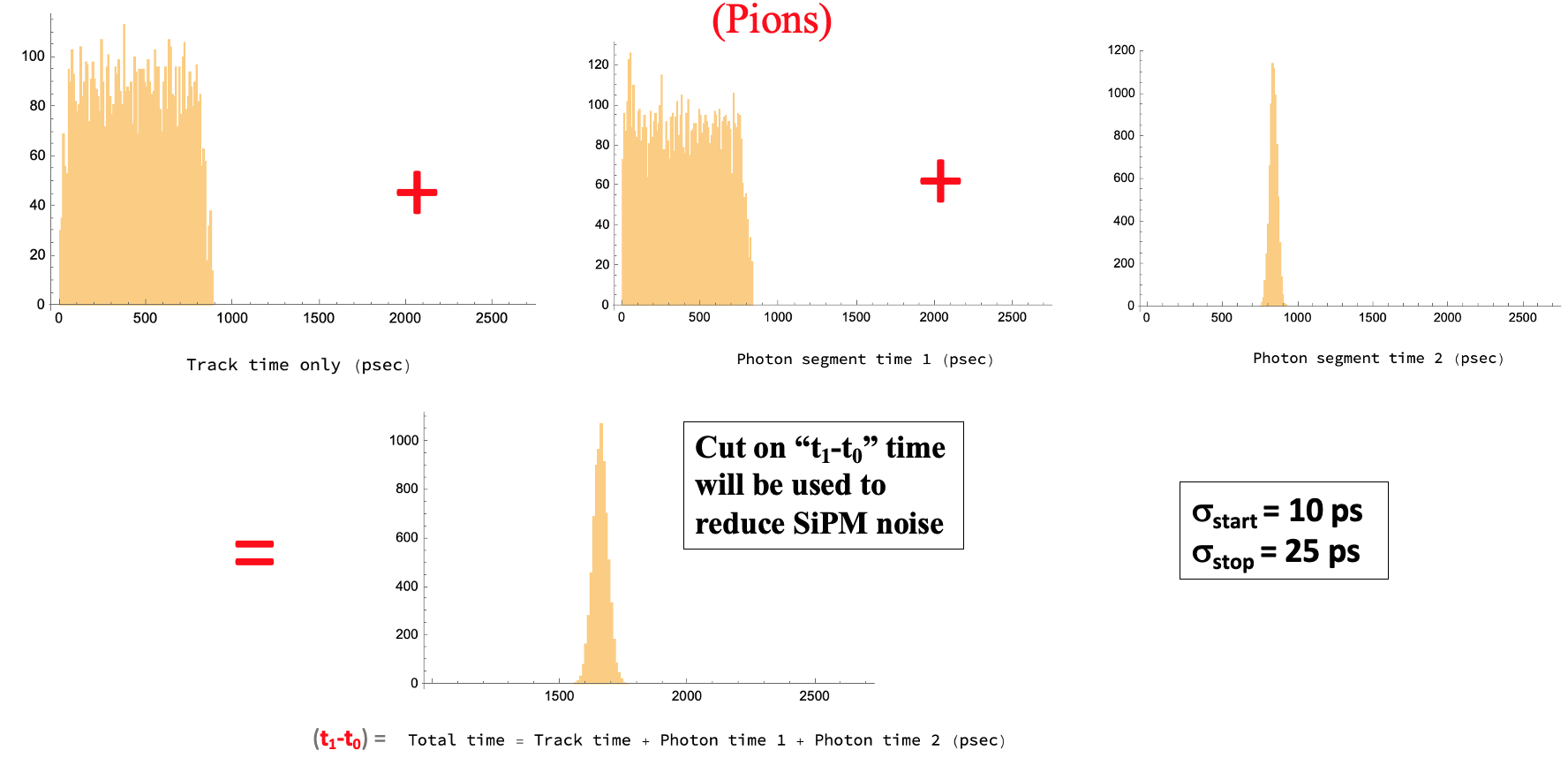}
    \caption{Time information for our proposed RICH for $\theta_\textrm{dip} = 4^\circ$, $p = \unit[20]{\GeVc}$, and $B = \unit[5]{T}$. One can see that $(t_1 - t_0)$ is a narrow distribution. We have chosen start and stop time resolutions of 10 and \unit[25]{ps}, respectively, with the former provided by a special timing layer in the overall detector.}
    \label{fig:RICH_times}
\end{figure}

\begin{figure}[htbp]
    \centering
    \includegraphics[width=\textwidth]{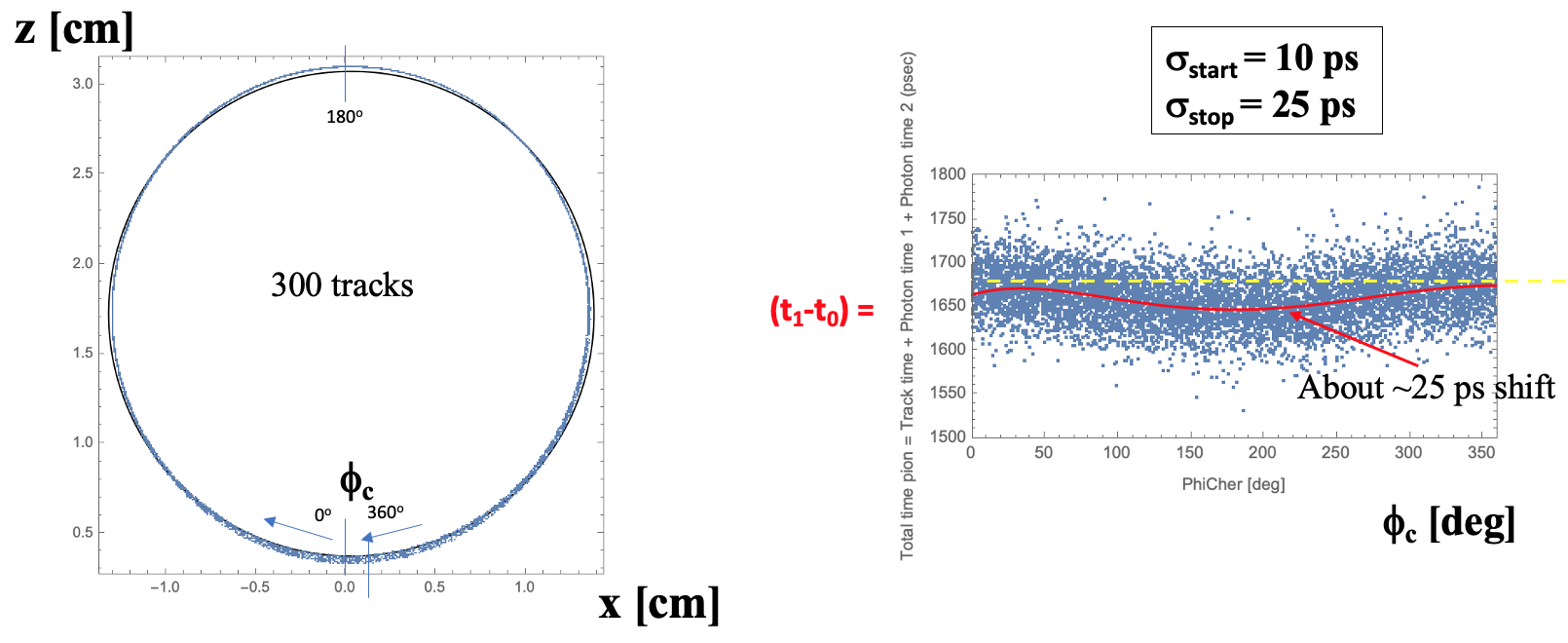}
    \caption{(Left) an image of a Cherenkov ring with the direction of the Cherenkov angle azimuth, $\phi_c$, indicated and (right) the dependency of $(t_1 - t_0)$ on $\phi_c$ for $\theta_\textrm{dip} = 4^\circ$, $p = \unit[20]{\GeVc}$, and $B = \unit[5]{T}$.}
    \label{fig:RICH_t1_t0}
\end{figure}

As schematically shown in Fig.~\ref{fig:simulation}, the program keeps track of time, namely (a) the track time before radiating a photon, (b) the photon time before reflection from the mirror, and (c) the photon time of reflected photon before it hits the SiPM plane. Fig.~\ref{fig:RICH_times} shows the various time distributions as well as their sum, which is a narrow distribution equal to $(t_1 - t_0)$, the difference between the photon stop hit and the track start hit. We plan to apply a cut on this time to reduce SiPM noise. Fig.~\ref{fig:RICH_t1_t0} shows the dependency of $(t_1 - t_0)$ on the Cherenkov angle azimuth $\phi_c$ for $\theta_\textrm{dip} = 4^\circ$, $p = \unit[20]{\GeVc}$, and $B = \unit[5]{T}$. The simulation indicates that the total time shift along the azimuth is about \unit[25]{ps}. However, it would be visible only for exceptional timing performance.

\begin{figure}[htbp]
    \centering
    \includegraphics[width=\textwidth]{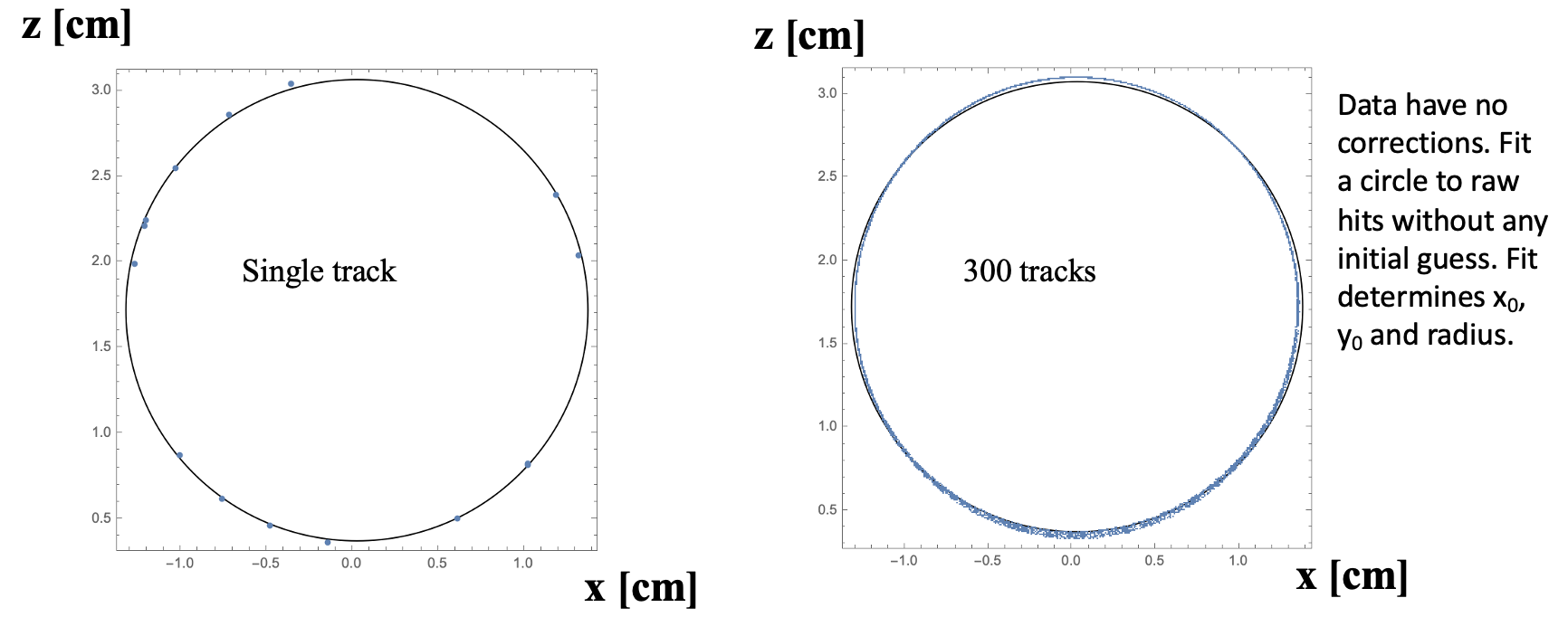}
    \caption{Cherenkov rings for $\theta_\textrm{dip} = 4^\circ$, $p = \unit[20]{\GeVc}$, and $B = \unit[5]{T}$ for (left) one track and (right) 300~tracks.}
    \label{fig:RICH_distortion}
\end{figure}

\begin{figure}[htbp]
    \centering
    \includegraphics[width=0.85\textwidth]{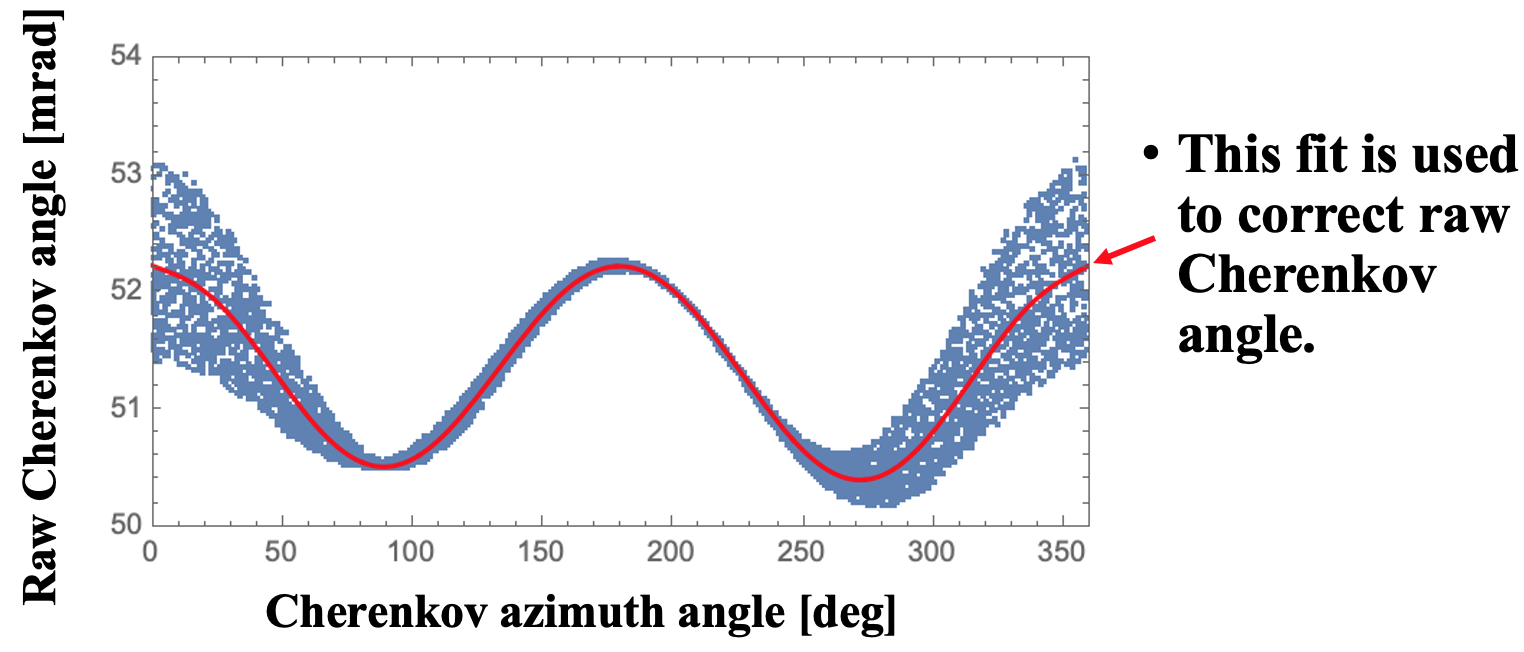}
    \caption{Raw Cherenkov angle as a function of Cherenkov azimuth for $\theta_\textrm{dip} = 4^\circ$, $p = \unit[50]{\GeVc}$, and $B = \unit[5]{T}$. The correction for the elliptical distortion of the Cherenkov ring is indicated by the red line.}
    \label{fig:RICH_residuals}
\end{figure}

Fig.~\ref{fig:RICH_distortion} shows a single Cherenkov ring for $\theta_\textrm{dip} = 4^\circ$, $p = \unit[20]{\GeVc}$, and $B = \unit[5]{T}$. Based on one event, one does not recognize any distortion; however, the distortion becomes clear in a sample of 300~tracks. One can see that the final image is actually an ellipse with out-of-focus regions at certain azimuths. We would therefore expect that the residuals relative to a circle will follow a sine wave. Indeed, that is what we see in Fig.~\ref{fig:RICH_residuals}, where we plot the raw Cherenkov angle, given by the Cherenkov radius divided by the focal length, as a function of the Cherenkov angle azimuth. Here, the Cherenkov radius for detector hit $i$ is given by $\sqrt{(x_\textrm{final}[i] - x_0)^2 + (z_\textrm{final}[i] - z_0)^2}$ where $x_0$ and $z_0$ correspond to the center of the circle, and the focal length is given by $R/2$ where $R$ is the radius of the spherical mirror. The fit of this dependency is shown in Fig.~\ref{fig:RICH_residuals}. We use this fit to correct for the elliptical distortion, yielding a corrected Cherenkov angle. This particular calculation was done for $\theta_\textrm{dip} = 4^\circ$, $p = \unit[50]{\GeVc}$, and $B = \unit[5]{T}$. One generally expects that this correction is dependent on the momentum and dip angle, and it corresponds to a \emph{major} improvement in obtaining a good Cherenkov angle resolution.

\begin{figure}[htbp]
    \centering
    \includegraphics[width=\textwidth]{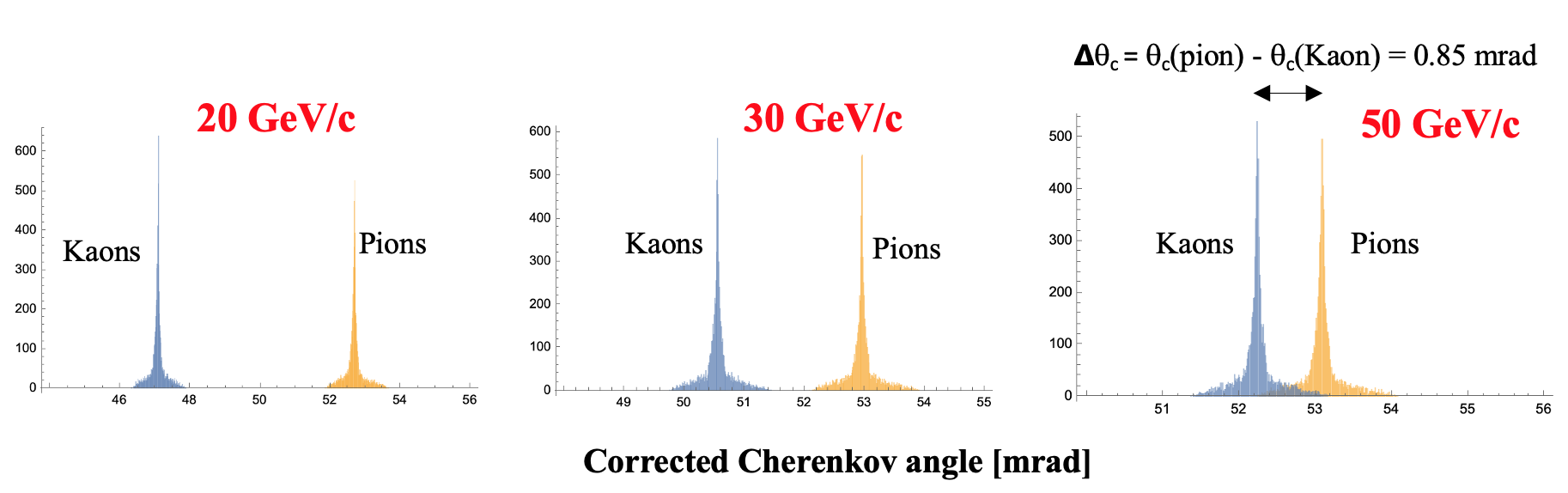}
    \caption{The corrected Cherenkov angle distribution for $\theta_\textrm{dip} = 4^\circ$ and $B = \unit[5]{T}$ and varying momenta.}
    \label{fig:RICH_corrected_4o}
\end{figure}

\begin{figure}[htbp]
    \centering
    \includegraphics[width=\textwidth]{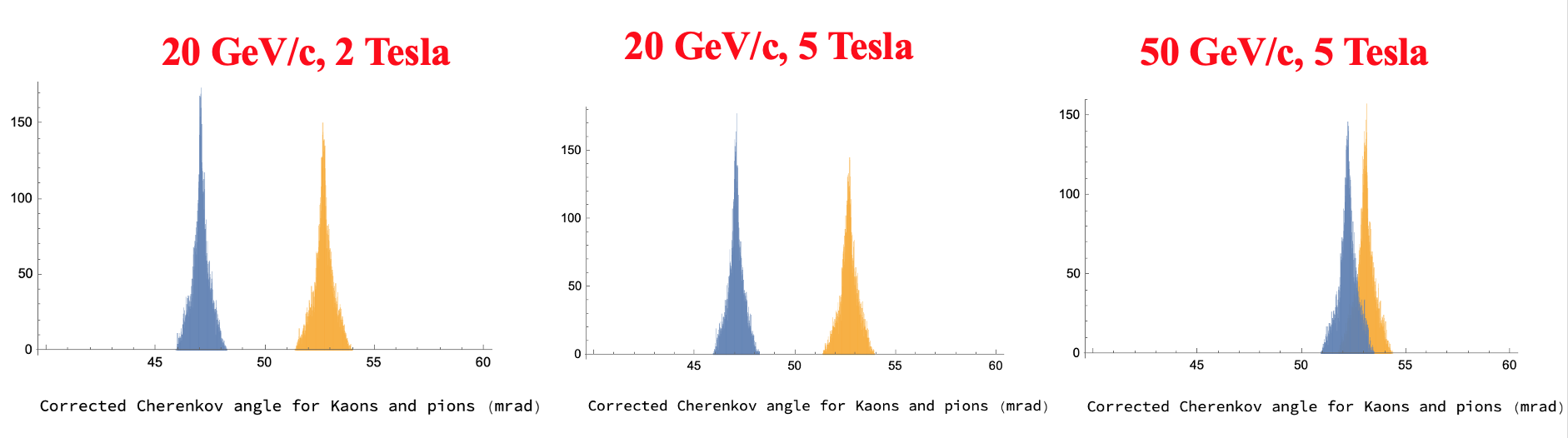}
    \caption{The corrected Cherenkov angle distribution for $\theta_\textrm{dip} = 40^\circ$ and varying momenta and magnetic fields.}
    \label{fig:RICH_corrected_40o}
\end{figure}

Fig.~\ref{fig:RICH_corrected_4o} shows corrected Cherenkov angle distributions, which only include focusing and smearing effects, for $\theta_\textrm{dip} = 4^\circ$ and $B = \unit[5]{T}$. By including the correction of Fig.~\ref{fig:RICH_residuals}, the Cherenkov angle distributions improve dramatically. The typical RMS error is \unit[$\sim$0.25]{mrad} per photon (includes tails).\footnote{It should be noted that because the distributions are not Gaussian, the RMS error does not correspond to the width of a Gaussian.} Fig.~\ref{fig:RICH_corrected_40o} shows the same distributions for $\theta_\textrm{dip} = 40^\circ$ and $B = 2$ or \unit[5]{T}. In this case, the typical RMS is \unit[$\sim$0.43]{mrad} per photon (includes tails). Comparing Figs.~\ref{fig:RICH_corrected_4o} and \ref{fig:RICH_corrected_40o}, we see that larger dip angles have larger RMS errors.

\begin{figure}[htbp]
    \centering
    \includegraphics[width=\textwidth]{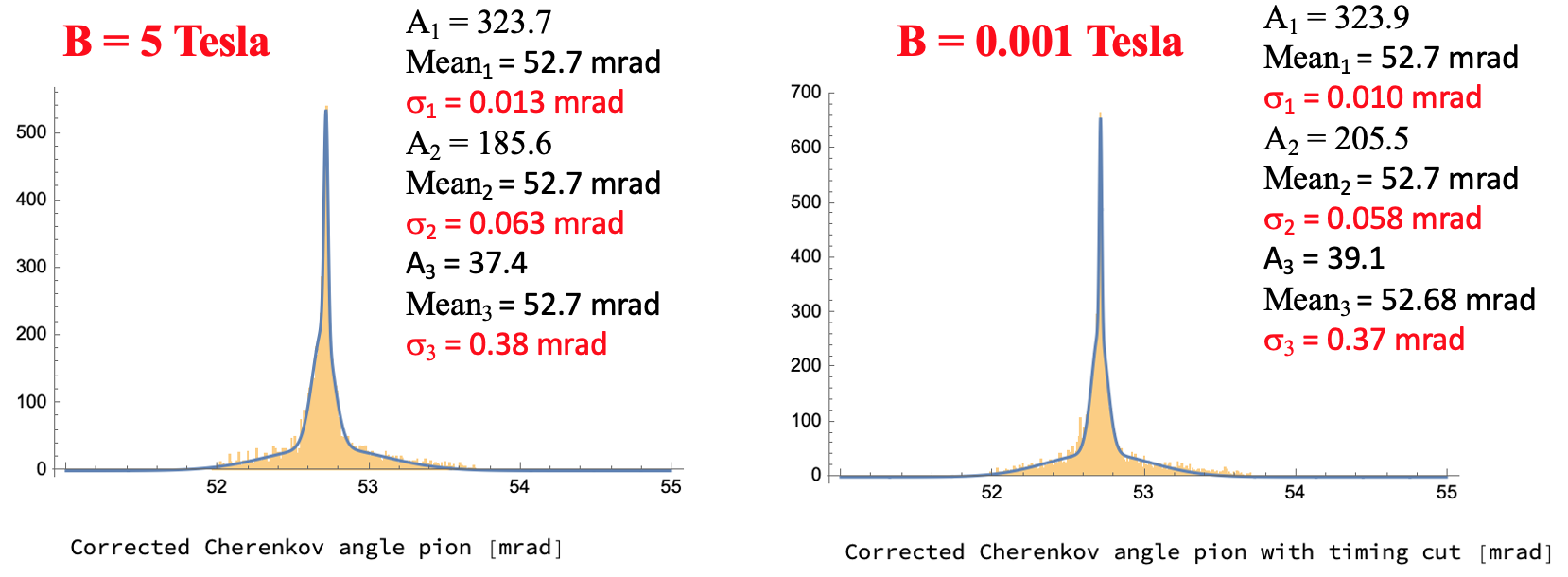}
    \caption{Cherenkov angle distributions --- including focusing and smearing effects --- as well as their corresponding fits for $\theta_\textrm{dip} = 4^\circ$ and $p = \unit[20]{\GeVc}$, with (left) and without (right) magnetic fields.}
    \label{fig:RICH_fits}
\end{figure}

Fig.~\ref{fig:RICH_fits} shows a fit to the Cherenkov angle resolution --- consisting of the product of three Gaussians --- for pions at \unit[20]{\GeVc} and $\theta_\textrm{dip} = 4^\circ$ as well as a magnetic field of either 5 or \unit[0.001]{T}. We conclude that for this momentum and detector orientation, the focusing error is larger than the smearing error. This is quantified in Table~\ref{tab:RICH_smearing_and_focusing}, where we have determined the smearing errors by subtracting the fitted error values shown in Fig.~\ref{fig:RICH_fits} in quadrature.

\FloatBarrier

\begin{table}[htbp]
    \centering
    \caption{Smearing and focusing errors for $\theta_\textrm{dip} = 4^\circ$ and $B = \unit[5]{T}$, based on fits to the Cherenkov angle distributions shown in Fig.~\ref{fig:RICH_fits}. The left column is calculated by subtracting in quadrature the fitted width obtained without a magnetic field from that obtained with a magnetic field.}
    \label{tab:RICH_smearing_and_focusing}
    \begin{tabular}{cc}
        \toprule
        Smearing error & Focusing error \\
        \midrule
        $\sigma_1 = \unit[0.008]{mrad}$ & $\sigma_1 = \unit[0.010]{mrad}$ \\
        $\sigma_2 = \unit[0.024]{mrad}$ & $\sigma_2 = \unit[0.058]{mrad}$ \\
        $\sigma_3 = \unit[0.090]{mrad}$ & $\sigma_3 = \unit[0.370]{mrad}$ \\
        \bottomrule
    \end{tabular}
\end{table}

\subsection{Total Cherenkov angle resolution, including all errors}

Table~\ref{tab:RICH_errors} shows the contributions to the final Cherenkov angle error for several dip angles and particle momenta, neglecting systematics errors. The final error is calculated assuming tracking error contributions of \unit[0.3]{mrad} as well as \unit[0.1]{mrad}. For the latter --- a very optimistic case --- one can see that the $\pi$/$K$ separation at \unit[50]{\GeVc} will exceed $4\sigma$ for $\theta_\textrm{dip} = 4^\circ$ and $4.6\sigma$ for $\theta_\textrm{dip} = 40^\circ$. Incidentally, the SiD detector’s tracking error may approach a value of \unit[$\sim$0.1]{mrad};\footnote{Private communication with the SiD group.} therefore, our compact RICH is a good match for SiD.

\begin{table}[htbp]
    \centering
    \caption{Contributions to the final Cherenkov angle error for the FBK SiPM design with a \unit[0.3]{mrad} tracking error. Overall standard deviation errors are used for all simulated distributions. The bracketed quantities correspond to the numbers obtained by assuming a \unit[0.1]{mrad} tracking error.}
    \label{tab:RICH_errors}
    \resizebox{\textwidth}{!}{
        \begin{tabular}{l|c|c|c|c}
            \toprule
            Momentum [$\GeVc$] & 20 & 30 & 50 & 50 \\
            $\theta_\textrm{dip}$ [$^\circ$] & 4 & 4 & 4 & 40 \\
            \midrule
            $N_\textrm{pe}$ per track for pions & 18 & 18 & 18 & 24 \\
            Chromatic error per photon hit [mrad] & 0.62 & 0.62 & 0.62 & 0.62 \\
            Chromatic error per track [mrad] & 0.143 & 0.143 & 0.143 & 0.125 \\
            \unit[0.5]{mm} pixel error per photon hit [mrad] & 0.38 & 0.38 & 0.38 & 0.38 \\
            \unit[0.5]{mm} pixel error per track [mrad] & 0.09 & 0.09 & 0.09 & 0.06 \\
            Focusing/smearing error per photon hit [mrad] & 0.25 & 0.25 & 0.25 & 0.44 \\
            Focusing/smearing error per track after correction [mrad] & 0.058 & 0.057 & 0.057 & 0.089 \\
            Track error [mrad] & 0.3 (0.1) & 0.3 (0.1) & 0.3 (0.1) & 0.3 (0.1) \\
            \midrule
            Total error per track [mrad] & 0.35 (0.21) & 0.35 (0.21) & 0.35 (0.21) & 0.34 (0.18) \\
            PID $\pi$/$K$ separation [number of sigma] & 16.0 (26.5) & 6.9 (11.4) & 2.4 (4.0) & 2.5 (4.6) \\
            \bottomrule
        \end{tabular}
    }
\end{table}

\begin{figure}[htbp]
    \centering
    \includegraphics[width=\textwidth]{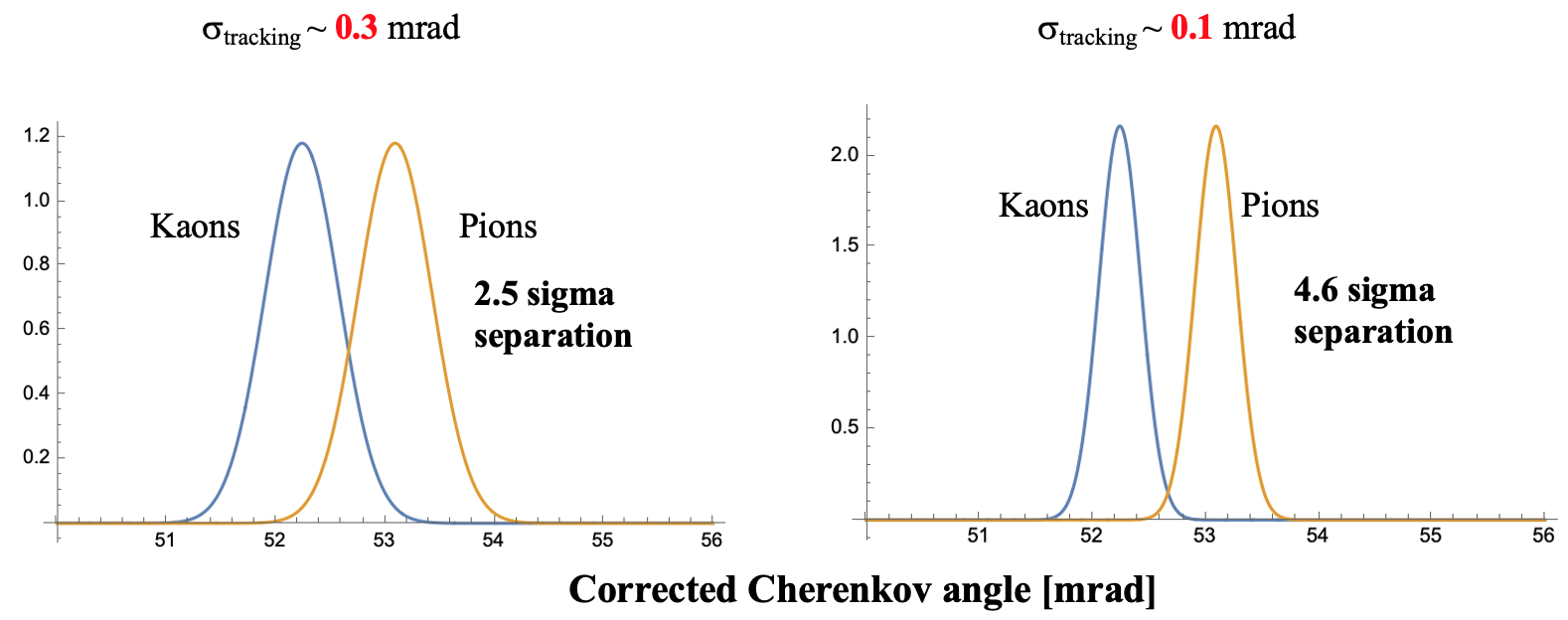}
    \caption{Expected final PID for $\theta_\textrm{dip} = 40^\circ$, $p = \unit[50]{\GeVc}$, and $B = \unit[5]{T}$ for two different tracking errors. In this plot, we consider all contributions to the final error from Table~\ref{tab:RICH_errors}.}
    \label{fig:RICH_final_PID}
\end{figure}

Fig.~\ref{fig:RICH_final_PID} shows the $\pi$/$K$ separation, considering all the error contributions of Table~\ref{tab:RICH_errors}, for $\theta_\textrm{dip} = 40^\circ$, tracking errors of 0.3 or \unit[0.1]{mrad}, and a track momentum of \unit[50]{\GeVc}.

\subsection{SiPM noise}
\label{sec:SiPM_noise}

\begin{figure}[htbp]
    \centering
    \begin{subfigure}[b]{0.49\textwidth}
        \centering
        \includegraphics[width=1.\textwidth]{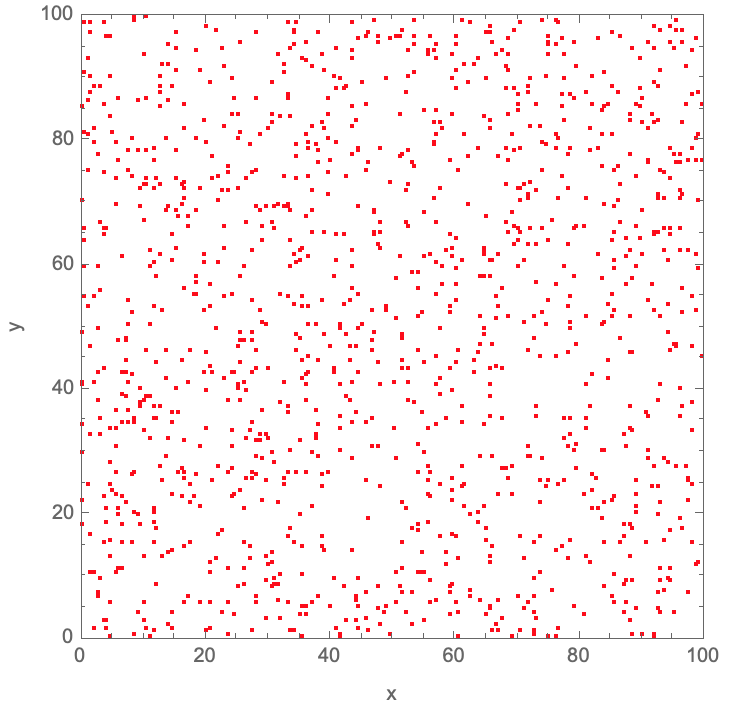}
        \caption{}
        \label{fig:SiPM_noise_no_cut}
    \end{subfigure}
    \hfill
    \begin{subfigure}[b]{0.49\textwidth}
        \centering
        \includegraphics[width=1.\textwidth]{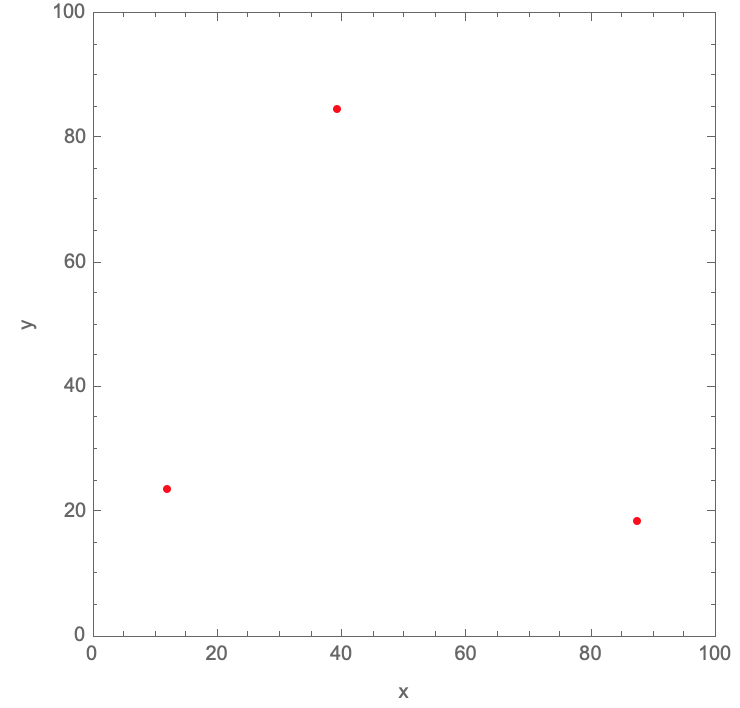}
        \caption{}
        \label{fig:SiPM_noise_with_cut}
    \end{subfigure} \\
    \caption{(a) The expected noise rate in an array $\unit[100]{mm}\times\unit[100]{mm}$ in size with a large \unit[100]{ns} window at room temperature. (b) The same array but with a \unit[$\pm$200]{ps} timing cut around the expected value of $(t_1 - t_0)$.}
    \label{fig:SiPM_noise}
\end{figure}

SiPM noise is important variable to consider --- the question is how SiPM random noise affects the RICH's operation. Usually, people try to solve this problem by cooling SiPMs to \unit[$-$30]{\oC} --- see \ref{app:SiPM_noise}. This cannot be done in our case as the C$_4$F$_{10}$ gas will liquefy at \unit[$-$2]{\oC}. We propose to cool the SiPMs to only \unit[+2--3]{\oC}. FBK says that the noise is reduced by a factor of 2 for every \unit[10]{\oC} drop in temperature; in other words, we can only hope to get a reduction by a factor of 4~\cite{Acerbi}. The remaining noise will be reduced by the timing cut. We were told by FBK that the noise rate at high PDE values could be as high as \unit[$\sim$100]{kHz} per ($\unit[0.5]{mm}\times\unit[0.5]{mm}$) pixel at room temperature~\cite{Acerbi}. Fig.~\ref{fig:SiPM_noise_no_cut} shows the noise rate at room temperature in an array $\unit[100]{mm}\times\unit[100]{mm}$ in size (40,000~pixels, each one running at \unit[100]{kHz}) with a crude \unit[100]{ns} timing cut, and Fig.~\ref{fig:SiPM_noise_with_cut} shows the same array but with a \unit[$\pm$200]{ps} timing cut around the expected value of $(t_1 - t_0)$ --- see Fig.~\ref{fig:RICH_times}. It is clear that timing helps a great deal. In addition to the timing cut, we get an additional factor of 4 reduction in the noise from the temperature decrease, virtually eliminating the SiPM noise. This calculation assumes that the physics and real background rates are small, which is satisfied at the next linear collider.

\subsection{Discussion of resolution study}

Table~\ref{tab:resolution} shows a summary of the various error contributions to the Cherenkov angle resolution. The SiD/ILD RICH design is compared with the SLD CRID gaseous RICH design. The SLD CRID had a local Cherenkov angle resolution per photon of \unit[$\sim$3.8]{mrad}, determined by fitting rings alone; however, the final overall resolution was quoted at a level of \unit[$\sim$4.3]{mrad} due to additional errors, yielding a resolution per track of $\unit[4.3/\sqrt{10}]{mrad} \approx \unit[1.3]{mrad}$. In addition, there were systematic errors including the (a) angular track resolution, (b) electron path and drift velocity in the TPC, (c) TPC position and orientation, (d) mirror position, orientation and radius, (e) refractive index variation due to radiator gas stability (i.e., mix and pressure), and (f) electronics gain. These effects made the CRID analysis difficult but successful~\cite{SLD:1998} --- see \ref{app:SLD_CRID}. Another thing to consider is cleanliness of tracks: the SLD CRID had $\sim$10~photoelectrons per ring for clean dimuon tracks; however, for tracks within jets, this number decreased to $\sim$8~photoelectrons per ring~\cite{Pavel:1997ds}.

The CRID error due to focusing/smearing was very small; in contrast, the SiD/ILD RICH has a larger chromatic error and a larger smearing/focusing error. To reduce the chromatic error, we can use filters in a form of micro-lenses to take care of SiPM edge effects, as suggested by Gola~\cite{GolaConf}. The focusing/smearing error at \unit[5]{T} could be reduced by tweaking the SiPM plane orientation; however, we found that the smearing effect is small above \unit[20]{\GeVc}. Below that momentum, it can be large, but particle separation is also larger. We cannot have a pixel size larger than $\unit[0.5]{mm}\times\unit[0.5]{mm}$, as it begins to dominate. Another critical contribution is the tracking angular resolution, which needs to be below \unit[0.3]{mrad} if one wants to achieve PID at \unit[50]{\GeVc}. The SiD tracking resolution is supposed to be \unit[0.1]{mrad}, which would be ideal for this type of RICH. For comparison, the SLD drift chamber provided the CRID with a tracking angular resolution of \unit[$\sim$0.8]{mrad}~\cite{Markiewicz, Hildreth:1995}. Many of the other systematic effects will not exist in our RICH design thanks to its solid-state photodetector choice. However, some resolution effects will remain similar, such as items (a), (d), (e) and (f) from the previous paragraph.

Table~\ref{tab:resolution} and Fig.~\ref{fig:expected_sigmas} also show the predicted PID performance for our RICH and the SLD CRID designs. The only way to improve this performance is to increase the gas pressure and to reduce the radial length, as shown in Ref.~\cite{FortyConf}. However, the price for this improvement is significant: one needs to deal with a pressure vessel holding \unit[3.5]{bar} and the increase in detector mass ($X/X_0 \sim 10\%$). With the help of a light mass vessel and the use of beryllium mirrors, the aim of the design is to keep material budget to $X/X_0 \sim \textrm{4--5\%}$.

\begin{table}[htbp]
    \centering
    \caption{Summary of the major error contributions of our RICH and the SLD CRID designs.}
    \label{tab:resolution}
    \begin{tabular}{l|c|c}
        \toprule
        & SiD/ILD RICH detector & SLD CRID detector \\
        & @ \unit[5]{T} [mrad] & @ \unit[0.5]{T} [mrad] \\
        \midrule
        \multicolumn{3}{l}{Error source} \\
        \midrule
        Chromatic error/photon & $\sim$0.62 & $\sim$0.4 \\
        Pixel size error/photon & 0.4 & $\sim$1.5 \\
        Smearing/focusing error/photon & 0.25--0.44 & $\sim$0.025 \\
        Total single-photon error/track & 0.18--0.35 & $\sim$1.35 \\
        \midrule
        \multicolumn{3}{l}{Other critical variables} \\
        \midrule
        $N_\textrm{pe}$/ring for $\beta \sim 1$ & $\sim$18 (for FBK SiPM) & $\sim$10 (dimuons in gas) \\
        $X/X_0$ & 4--5\% & $>$15\% \\
        \bottomrule
    \end{tabular}
\end{table}

\section{Conclusion}
\label{sec:conclusion}

This simple study indicates that there is a hope for PID up to \unit[50]{\GeVc} using our RICH design at the SiD or ILD detectors operating at \unit[5]{T}. The final performance, shown in Table~\ref{tab:resolution} and Fig.~\ref{fig:expected_sigmas}, critically depends on the Cherenkov angle resolution.

We have demonstrated that we can deal with optical distortions resulting in elliptical rings and that a tight timing cut can significantly reduce the SiPM random noise. As the SLD CRID proved, it is possible to reach the ultimate goal~\cite{SLD:1998}. The SLD CRID was hard to build, commission, run, and analyze, but unprecedented performance was achieved~\cite{SLD:2003ogn}.

In terms of next steps, it is necessary to optimize the optical design of the entire system considering all tracks and all momenta. This requires finding the optimum orientation of the detectors and tuning the mirror parameters. Once we have a basic geometry defined, we can perform a full \GeantFour simulation of the entire system. Additionally, photon detectors such as SiPMs will likely improve significantly over next 10 years, and so many parameters chosen for this study will likely improve.

\FloatBarrier

\appendix

\section{Additional discussion on PID reach by various PID techniques}
\label{app:PID_reach}

\begin{figure}[htbp]
    \centering
    \includegraphics[width=0.7\textwidth]{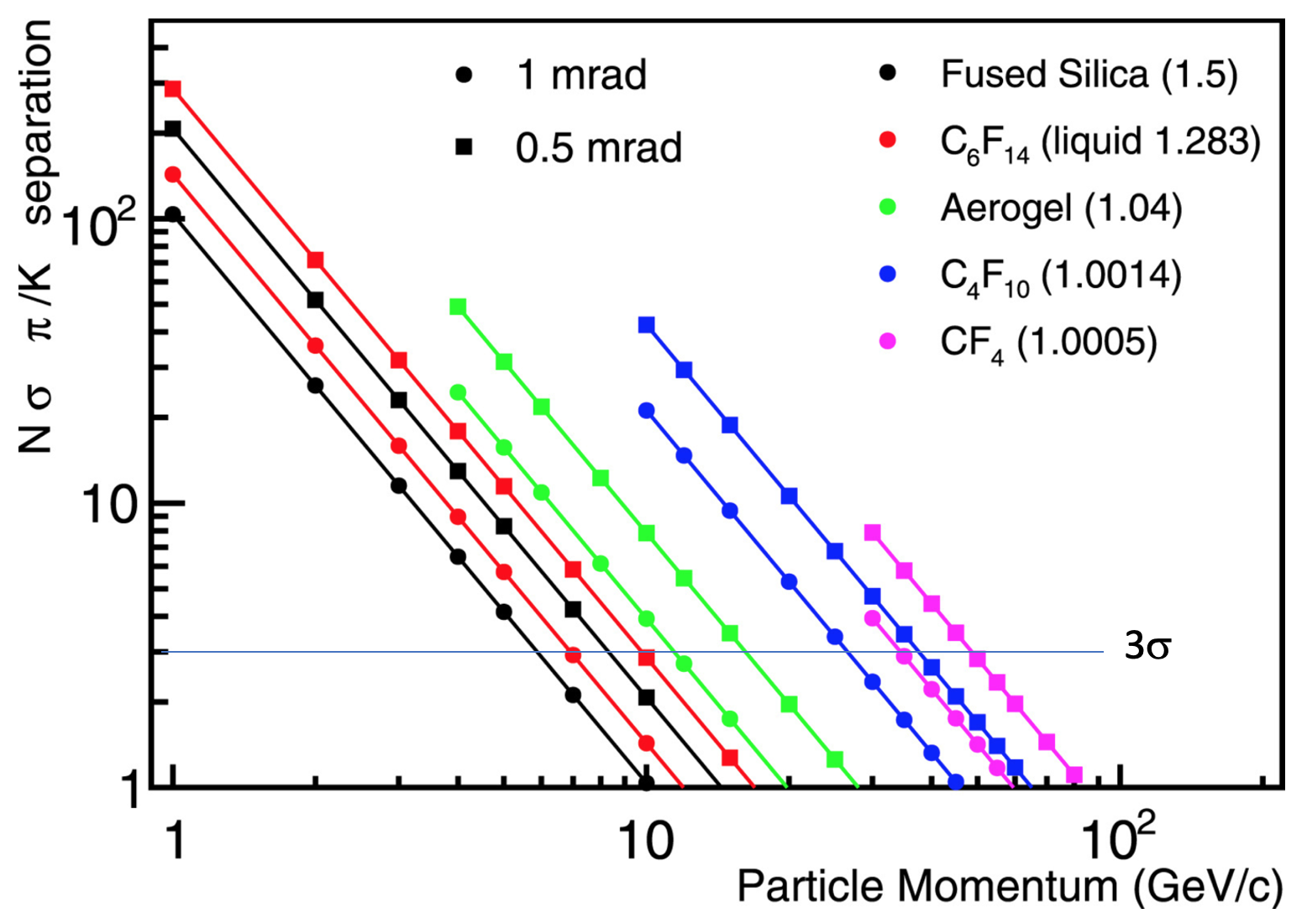}
    \caption{Expected $\pi$/$K$ separation reach in terms of number of sigma for various radiator choices and for two Cherenkov angle resolutions, $\sigma_\textrm{tot} = 0.5$ and \unit[1]{mrad}~\cite{Papanestis:2020, PapanestisConf}.}
    \label{fig:kaon_pion_separation}
\end{figure}

Figure~\ref{fig:kaon_pion_separation} shows the $\pi$/$K$ separation versus particle momentum for different radiators --- solid, liquid, and gaseous --- and two different values of the total Cherenkov angle resolution, $\sigma_\textrm{tot} = 0.5$ and \unit[1]{mrad}~\cite{Papanestis:2020, PapanestisConf}. In practice, the resolution tends to be worse when all contributions are included.

\begin{figure}[htbp]
    \centering
    \includegraphics[width=\textwidth]{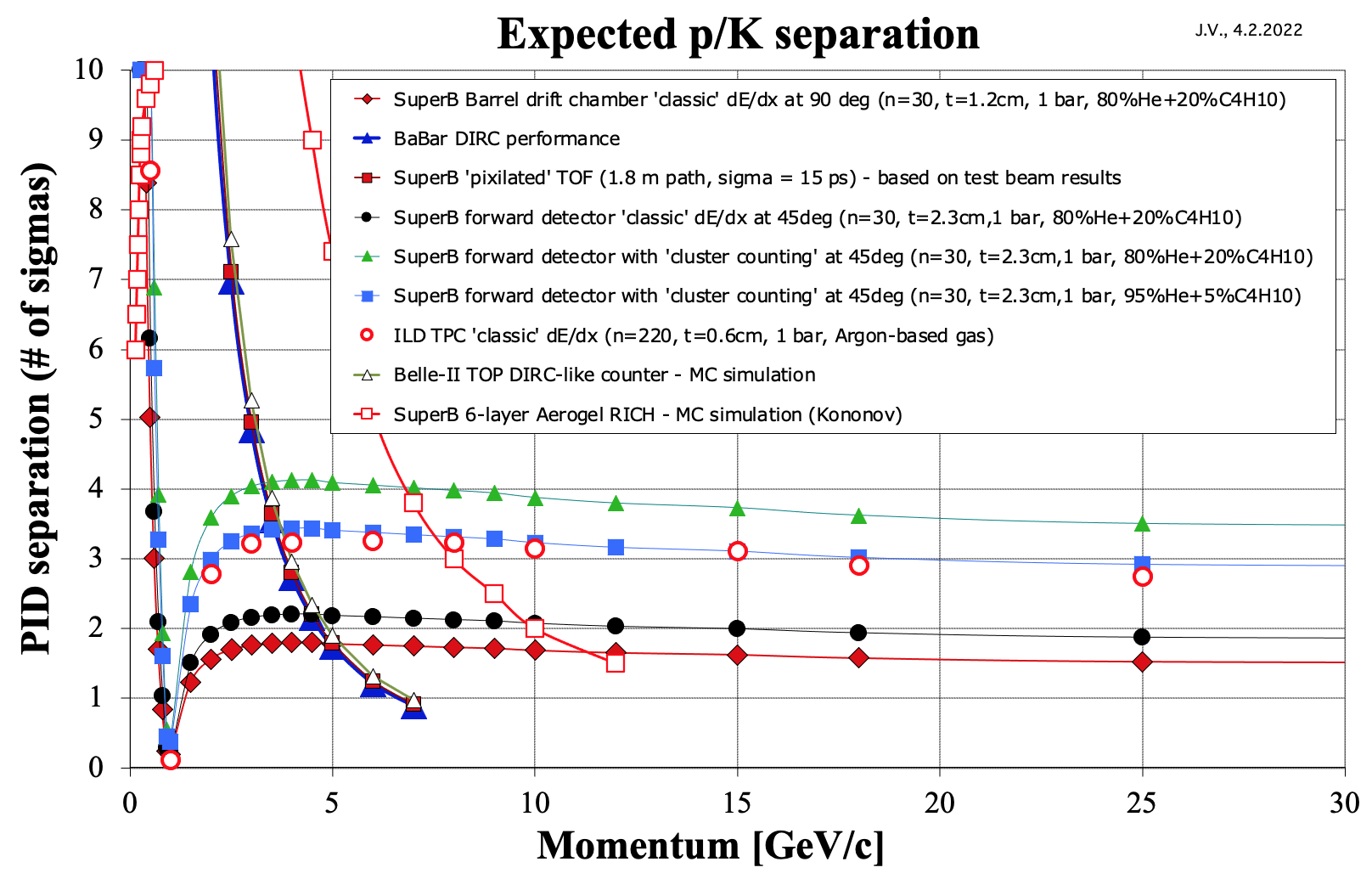}
    \caption{Expected $\pi$/$K$ separation~\cite{Vavra} reach in terms of number of sigma for TOF, $\dd{E}/\dd{x}$, BaBar DIRC~\cite{BaBarDIRC:2000smj}, Belle-II TOP counter~\cite{Nishimura:2010gj}, aerogel detectors within SuperB~\cite{SuperB:2013} and Belle-II~\cite{BelleII:2010}, and ILD TPC $\dd{E}/\dd{x}$ MC simulation~\cite{ILDConceptGroup:2020}.}
    \label{fig:lots_of_kaon_pion_separation}
\end{figure}

Fig.~\ref{fig:lots_of_kaon_pion_separation} shows the PID performance~\cite{Vavra} for a TOF counter with \unit[1.8]{m} flight path, the SuperB drift chamber $\dd{E}/\dd{x}$, the BaBar DIRC~\cite{BaBarDIRC:2000smj}, the Belle-II time-of-propagation (TOP) counter~\cite{Nishimura:2010gj}, aerogel detectors within SuperB~\cite{SuperB:2013} and Belle-II~\cite{BelleII:2010}, and the ILD TPC $\dd{E}/\dd{x}$ MC simulation~\cite{ILDConceptGroup:2020}. We see that the cluster counting method improves the PID when compared to classical $\dd{E}/\dd{x}$ in the SuperB drift chamber and that even a high performance TOF counter with a \unit[25]{ps} resolution has a limited reach up to only \unit[$\sim$3]{\GeVc}.

\FloatBarrier

\section{Additional discussion on SiPM noise}
\label{app:SiPM_noise}

\begin{figure}[htbp]
    \centering
    \includegraphics[width=\textwidth]{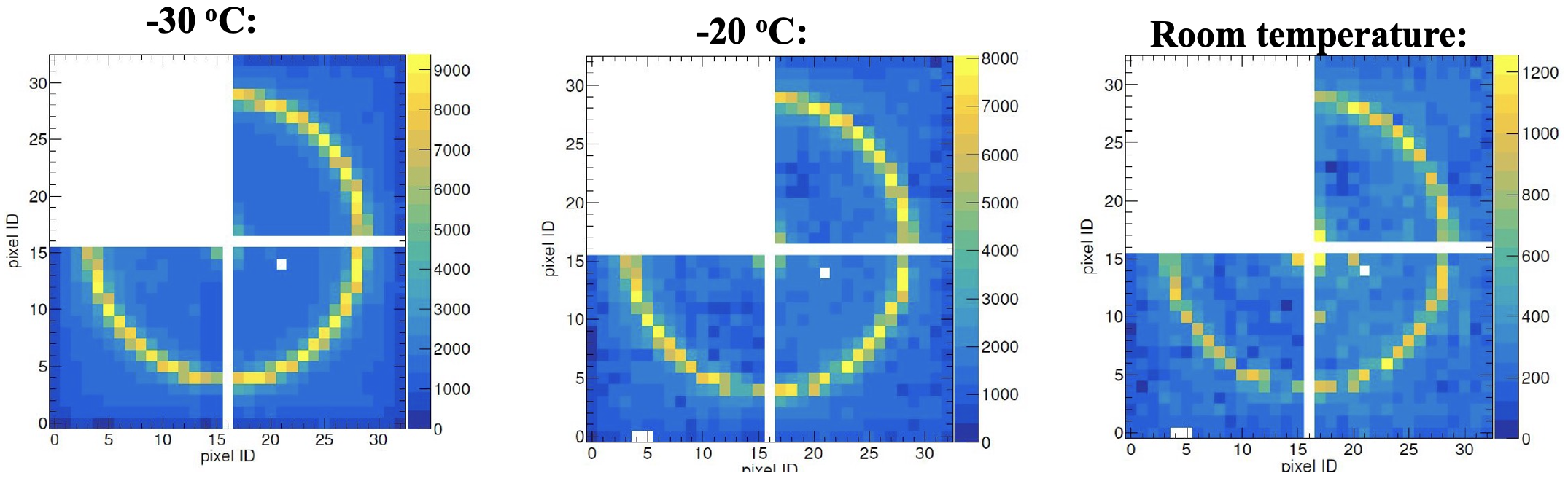}
    \caption{SiPM thermal random noise in images of Cherenkov rings as a function of temperature. These are results from EIC detector R\&D~\cite{HeAndSchwiening, Wong}.}
    \label{fig:SiPM_noise_temp}
\end{figure}

The main advantage of SiPMs is that they can certainly operate at \unit[5]{T} and even at \unit[7]{T}~\cite{Espana:2017}. However, compared to an ideal photon detector, the SiPM performance is affected by random dark noise~\cite{Klanner:2018}. It was an open question until a few years ago if they are suitable for a RICH imaging application. However, several experiments proved that the noise can be managed by lowering the SiPM temperature. Fig.~\ref{fig:SiPM_noise_temp} shows an example of an aerogel RICH detector being developed for an electron-ion collider (EIC) whose noise is controlled by temperature~\cite{HeAndSchwiening, Wong}. The noise gets worse if SiPMs are exposed to a total integrated neutron flux~\cite{Korpar:2020}; however, neutron backgrounds are predicted to be very low at SiD/ILD. One should note that this test did not employ a tight timing cut, as we have proposed in Section~\ref{sec:SiPM_noise}.

\FloatBarrier

\section{Physics performance of the SLD CRID}
\label{app:SLD_CRID}

\begin{figure}[htbp]
    \centering
    \begin{subfigure}[b]{0.55\textwidth}
        \centering
        \includegraphics[width=1.\textwidth,valign=b]{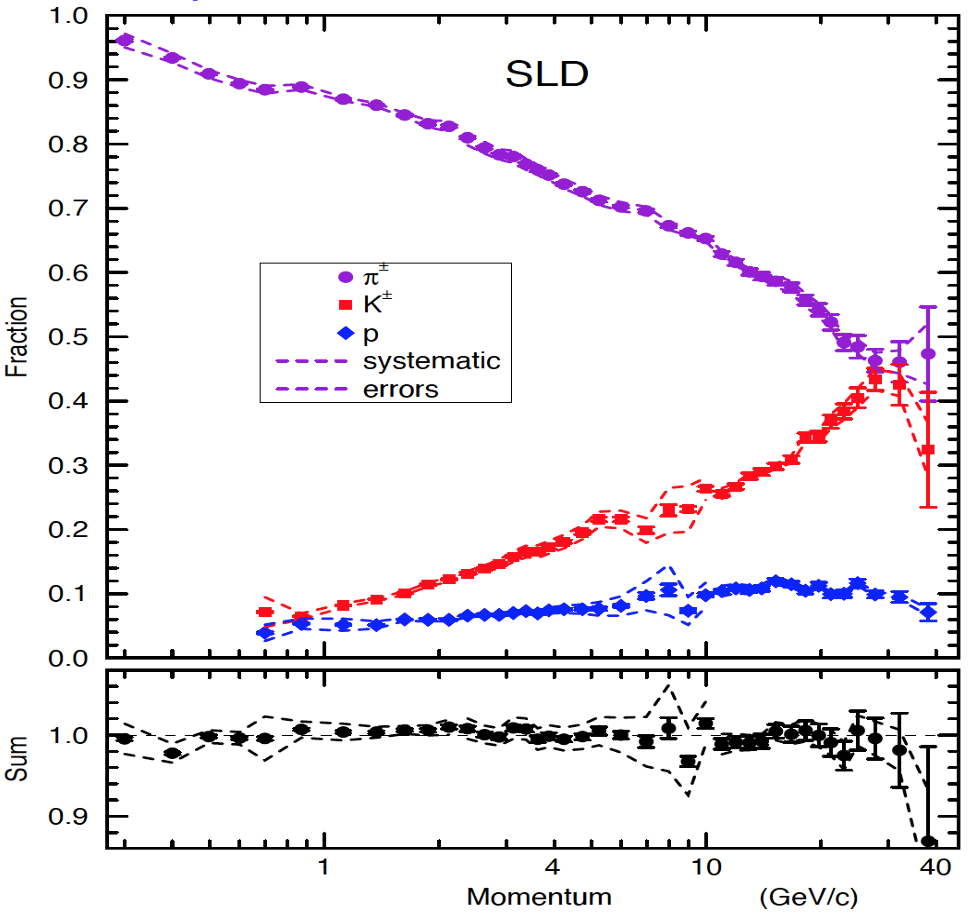}
        \caption{}
    \end{subfigure}
    \hfill
    \begin{subfigure}[b]{0.44\textwidth}
        \centering
        \includegraphics[width=1.\textwidth,valign=b]{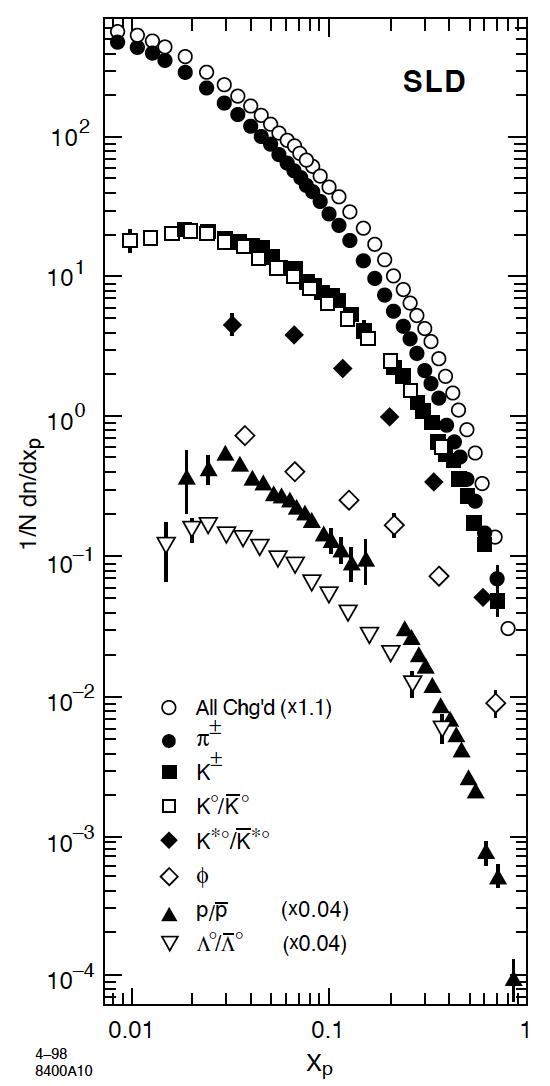}
        \caption{}
    \end{subfigure} \\
    \caption{(a) $\pi$/$K$/$p$ fractions determined by the SLD CRID~\cite{SLD:1998}. (b) Differential cross sections as a function of hadronic momentum fraction $x_p$ per hadronic $Z^0$ decay, by all SLD detectors~\cite{SLD:1998}.}
    \label{fig:SLD_CRID_piKp}
\end{figure}

Fig.~\ref{fig:SLD_CRID_piKp} demonstrates the physics achieved with a \unit[4.3]{mrad/photon} (or \unit[1.3]{mrad/track}) Cherenkov angle resolution at the SLD CRID~\cite{SLD:1998}. The PID limit for $\pi$/$K$ separation is between 25 and \unit[30]{\GeVc}.

\FloatBarrier

\addcontentsline{toc}{section}{References}
\bibliographystyle{elsarticle-num}
\bibliography{bibliography}

\end{document}